\documentstyle[eqsecnum,aps,epsf]{revtex}

\newcommand{\citen}{\cite} 
\newcommand{\BQ}{\begin{equation}}
\newcommand{\EQ}{\end{equation}}
\newcommand{\BQA}{\begin{eqnarray}}
\newcommand{\EQA}{\end{eqnarray}}

\newcommand{\NN}{\nonumber \\}

\newcommand{\Gmu}{\gamma^{\mu}}

\newcommand{\gfive}{\gamma_5}
\newcommand{\del}{\partial}
\newcommand{\k}{\mbox{\boldmath $k$}}

\newcommand{\itP}{\mbox{\boldmath $P$}}

\newcommand{\p}{\mbox{\boldmath $p$}}
\newcommand{\x}{\mbox{\boldmath $x$}}
\newcommand{\y}{\mbox{\boldmath $y$}}
\newcommand{\Z}{{\bf Z}}

\newcommand{\M}{{\cal M}}
\newcommand{\T}{{\cal T}}
\newcommand{\U}{{\cal U}}

\newcommand{\ket}[1]{\left.\left\vert #1 \right. \right\rangle}
\newcommand{\bra}[1]{\left\langle\left. #1 \right\vert\right.}
\newcommand{\ketrm}[1]{\vert {\rm #1} \rangle}  
\newcommand{\zm}[1]{\stackrel{\circ} {#1} }

\begin{document}

\topmargin=-1.5cm

\title{Light-Front Realization of Chiral Symmetry Breaking}

\author{Kazunori Itakura\footnote{\tt itakura@bnl.gov}}
\address{RIKEN-BNL Research Center, Brookhaven National Laboratory,
Upton, NY 11973, USA}

\author{Shinji Maedan\footnote{\tt maedan@tokyo-ct.ac.jp}}
\address{Department of Physics, Tokyo National College of Technology,
 Tokyo 193-0997, Japan}

\date{February 26, 2001}
\maketitle

\begin{abstract}
We discuss a description of chiral symmetry breaking in the light-front (LF)
  formalism. 
Based on careful analyses of several models, we give clear answers to 
  the following three fundamental questions: 
  (i) What is the difference between the LF chiral transformation and 
      the ordinary chiral transformation?
 (ii) How does a gap equation for the chiral condensate emerge? 
(iii) What is the consequence of the coexistence of a nonzero
       chiral condensate 
      and the trivial Fock vacuum?
The answer to Question (i) is given through a classical analysis of
  each model.
Question (ii) is answered based on our recognition of the importance of 
  characteristic constraints, such as the zero-mode and fermionic constraints.
Question (iii) is intimately related to another important problem, 
  reconciliation of the nonzero chiral condensate 
  $\langle \bar\Psi\Psi\rangle\neq 0$ and the invariance of the vacuum 
  under the LF chiral transformation $Q_5^{\rm LF}\ket{0}=0$.
This and Question (iii) are understood in terms of the modified chiral 
  transformation laws of the dependent variables.
The characteristic ways in which the chiral symmetry breaking is realized
  are that the chiral charge $Q_5^{\rm LF}$ 
  is no longer conserved and that the transformation of the scalar and 
  pseudoscalar fields is modified.  
We also discuss other outcomes, such as the light-cone wave function 
  of the pseudoscalar meson in the Nambu-Jona-Lasinio model.\\
\end{abstract}

\begin{center}
\noindent{\large\bf CONTENTS}
\end{center}

\contentsline {section}{\numberline {I}Introduction}{2}
\contentsline {section}{\numberline {II}Chiral transformation on the light front}{5}
\contentsline {subsection}{\numberline {A}Definition}{5}
\contentsline {subsection}{\numberline {B}Massive free fermion}{6}
\contentsline {subsection}{\numberline {C}Scalar field}{7}
\contentsline {subsection}{\numberline {D}Order parameters}{8}
\contentsline {section}{\numberline {III}Appearance of the gap equations: Model analysis}{8}
\contentsline {subsection}{\numberline {A}The sigma model -- Scalar fields}{9}
\contentsline {subsection}{\numberline {B}The chiral Yukawa model -- Scalar and fermionic fields}{10}
\contentsline {subsection}{\numberline {C}The Nambu--Jona-Lasinio model -- Fermionic fields }{13}
\contentsline {subsubsection}{\numberline {1}Classical solution of the fermionic constraint}{13}
\contentsline {subsubsection}{\numberline {2}Quantum and nonperturbative solution}{14}
\contentsline {section}{\numberline {IV}Chiral symmetry breaking on the light front}{15}
\contentsline {subsection}{\numberline {A}Chiral condensate and problem of mass information loss}{15}
\contentsline {subsection}{\numberline {B}Multiple Hamiltonians}{17}
\contentsline {subsection}{\numberline {C}The NG bosons}{18}
\contentsline {subsection}{\numberline {D}Modified chiral transformation}{19}
\contentsline {subsection}{\numberline {E}Nonconservation of the LF chiral charge}{20}
\contentsline {section}{\numberline {V}Coupling to gauge fields}{21}
\contentsline {section}{\numberline {VI}Conclusions}{22}
\contentsline {section}{\numberline {APPENDIXES\hskip 0pt plus1fill minus1fill\relax }{}}{23}

\section{Introduction}

Dirac's first proposal of the ``front form,'' which is now called the
  light-front (LF) Hamiltonian formalism, intended to combine  
  special relativity and quantum mechanics.\cite{Dirac} \ 
Since the quantum mechanics is based on Hamiltonian formalism, 
  Dirac's implied intention was to find the most convenient ``form'' 
  for relativistic quantum systems. 
In the usual instant form, with the equal-time hypersurface 
  $x^0=$ const,\footnote{In what follows, we simply refer to this standard 
  scheme as the equal-time (ET) quantization.} \ 
  translations and rotations transform quantum states in a very simple 
  manner, because they do not change the quantization surface.
These kinds of static Poincar\'e generators are called ``kinematical''.
The other generators, which change the quantization 
  surface, are called ``dynamical''.
For example, the Lorentz boost mixes time and space coordinates and 
  thus changes the hypersurface.
The Hamiltonian itself, of course, is a dynamical operator.
Therefore, an eigenstate in a rest frame is no longer an eigenstate 
  in the boosted frame. 
Even though we know eigenstates in the rest frame, 
  to find a new eigenstate in the boosted frame requires, in principle,
  as much as effort as solving the entire problem.
Contrastingly, the front form in which we treat $x^+=(x^0+x^3)/\sqrt2$
  as time,  allows the maximum number of kinematical operators 
  in ten Poincar\'e generators.
In particular, the boost operator now forms a part of the {\it kinematical} 
  operators.
With the LF coordinates,\footnote{Our definition of the 
  light-front variables and some conventions are given in Appendix A.} \  
  the boost transformation along the third axis is simply given by 
  $x^\pm \to {\rm e}^{\pm\phi} x^\pm\ (\tanh \phi=\beta)$, and it
  does not change the quantization surface $x^+=0$.
Due to this simplicity, it is quite easy to construct a boosted eigenstate 
  $\ket{n; P'}$ from $\ket{n; P}$, where $P'^\mu=\Lambda^\mu{}_\nu P^\nu$.
Therefore, the front form is considered to be an ideal framework for 
  the description of relativistic quantum systems.

This idea is convenient for understanding the hadron physics 
  based on the quantum chromodynamics (QCD), because all the hadrons are 
  relativistic quantum bound states of quarks and gluons.
Of course there exists a covariant formalism of quantum field theories, 
  but at present its success is, strictly speaking, limited only to the 
  perturbative region, and thus effort should be made to consider 
  other possible frameworks for relativistic quantum systems.
One of the merits of using Hamiltonian formalism is 
  that we can utilize various nonperturbative techniques 
  (such as the variational approximation, the Tamm-Dancoff approximation, etc.)
  developed in quantum mechanics.
Moreover, in some aspects, the LF field theory has a structure similar 
  to that of the nonrelativistic theory. 
This resemblance also allows us to use familiar nonperturbative methods
  of nonrelativistic quantum mechanics.
The other merits of the Hamiltonian formalism are that we can compute 
  both the wave functions and eigenvalues simultaneously, and that 
  it can in principle describe the time ($x^+$) evolution of the system. 
The accessibility to such dynamical information deserves more attention 
  and investigation.

Dirac's LF Hamiltonian formalism has been applied to 
  various quantum field theories.\cite{Review} \ 
The most significant success was made in 1+1 dimensional QCD, 
  where the mass spectra and wavefunctions of various mesons and baryons 
  for various numbers of colors were calculated by explicit diagonalization 
  of the LF Hamiltonian.\cite{Hornbostel} \  
An interesting application of the LF formalism to time dependence 
  of a relativistic system was made for the scattering 
  of two {\it composite} particles.\cite{Scatt}

Many of the properties of the LF formalism are related to the unique 
  structure of the dispersion relation
\BQ
p^-=\frac{p_\perp^2+m^2}{2p^+},\label{disp}
\EQ
where $p^\mp=(p^0\mp p^3)/\sqrt2$ are the LF energy ($-$) and the longitudinal 
 momentum (+).
Unlike the usual instant form dispersion $E=\pm\sqrt{\p^2+m^2}$, 
  the LF dispersion relation has a rational structure.\footnote{If we  
  regard $p^+$ as mass, Eq.~(\ref{disp}) can be understood as a 
  nonrelativistic dispersion in  (transverse) 2 dimensions, 
  $E_\perp=p_\perp^2/2m_\perp+E_0$.} \ 
This leads to an important property,  ``vacuum simplicity''.
Requiring the semi-positivity of the LF energy, $p^-\geq 0$, we immediately 
 find a restriction on the longitudinal momentum, $p^+\geq 0$. 
Therefore, no particle state with nonzero momentum can mix with 
 the Fock vacuum $\ket{0}$, which has zero momentum.
(We cannot construct a zero momentum from positive momenta.) 
This fact implies that the structure of the vacuum is very simple.
Note also that the large value of the LF energy involves two different cases,
  that of the small $p^+$ limit and that of the large $p_\perp$ limit.
If we set the infrared cutoff as $\epsilon<p^+$ to regularize the divergent 
  energy, the subtlety of the $p^+=0$ case is removed, and the vacuum of 
  the system becomes the trivial Fock vacuum.
This is one of the reasons why some people use the somewhat exaggerated 
  term  ``vacuum triviality''.

The LF vacuum also simplifies the description of excited states.
Roughly speaking, this is because any Fock state should be directly 
  related to the excited states.
Indeed, in a  method called the discretized light-cone quantization  
 (DLCQ) method,\cite{Maskawa-Yamawaki,Pauli-Brodsky}\ 
  which is discussed below, 
  the dimension of the Hamiltonian matrix in the longitudinal direction
  is finite in the calculation, 
  and the valence description is a good approximation for excited states.
This situation is very advantageous for QCD, since it suggests 
  the constituent picture, which allows us an intuitive understanding of 
  low energy hadronic properties.
In fact, one of the strong motivations which drives the present 
  resurgence of the LF field theory is our anticipation that it might connect 
  the constituent quark model and QCD at a field theoretic 
  level.\cite{Wilson} \

However, we must be careful not to be too optimistic about such a naive 
  expectation: The constituent picture is not just a valence picture. 
The constituent picture should emerge as a result of a significant
  nonperturbative phenomenon, the dynamical chiral symmetry breaking 
  (D$\chi$SB).\cite{Chiral}\ 
According to the conventional formulation,
  D$\chi$SB is thought to be described by determining a new vacuum state
  that breaks the chiral symmetry but is energetically favored.
A condition that minimizes the vacuum energy is the so-called gap equation, 
  which is generally a nonlinear equation for the order parameter of interest. 
On the other hand, the structure of the LF vacuum is kinematically determined 
  to be simple and seems independent of the dynamics.
So a natural question arises: 
{\it How can we reconcile a LF ``trivial'' 
  vacuum with a chirally broken ET vacuum having a nonzero fermion condensate?}
Actually this problem is not restricted to D$\chi$SB.
Most nonperturbative phenomena are currently believed to be related to 
  some nontrivial structure of the vacuum. 
In addition to D$\chi$SB, such ``vacuum physics'' may include
  topological effects (theta vacua), gluon condensation, the Higgs mechanism, 
  and even confinement.  
Therefore, we should ask more generally the following: 
  {\it How can we represent ``vacuum physics'' in the LF field theories?}
The primary purpose of the present paper is to determine a way of describing
  D$\chi$SB on the LF. 
Such consideration is necessary  for realizing the constituent 
  picture, but we hope that it will also provide 
  useful information for the understanding of other types of vacuum physics.

We wish to answer the above questions, and we now set out to do so.
It is now widely accepted that vacuum physics will be described on the LF 
  as  {\it physics of the  longitudinal zero mode} (simply referred to as 
  the ``zero mode''), which is the only point with respect to which 
  the vacuum triviality becomes subtle.
In other words, if the LF formalism is a correct framework, all the 
  information concerning vacuum physics should be extractable from 
  the vicinity of the zero mode.
This is a very delicate problem involving the infrared (IR) 
  singularity resulting from an infinitesimally small longitudinal momentum.
Therefore, the most important point when we discuss  vacuum 
  physics on the LF is determining  how to regularize the IR singularity.
Generally speaking,  three different methods have been proposed for the 
  study of such ``zero-mode physics''.
The relation among them is not clear at present time.
We discuss these methods below.\\

\vspace{-3mm}

\begin{description}
\item {1.} {\it The DLCQ method} \cite{Maskawa-Yamawaki,Pauli-Brodsky}\\
The most straightforward but powerful method is the DLCQ method, which treats 
  the zero mode explicitly by compactifying the longitudinal space into 
  a circle $x^-\in [-L, L]$ with appropriate boundary conditions on 
  fields.\cite{Maskawa-Yamawaki}\ 
All the longitudinal momenta take discrete values.
For a scalar field, we impose the periodic boundary condition at each LF 
  time $\phi(x^- =-L,~ x_\perp) = \phi(x^-= L,~ x_\perp)$, so that we can 
  unambiguously define the longitudinal zero mode: 
\BQ
\phi_0(x_\perp)=\frac{1}{2L}\int_{-L}^L dx^-\ \phi (x).
\EQ
Then the scalar field is decomposed into the zero mode and the 
  remaining oscillation modes: $\phi(x)=\phi_0(x_\perp)+\varphi(x).$
Since the vacuum expectation value of $\varphi$ is zero, the central 
  issue is determining how to evaluate the zero-mode part 
  $\bra{0}\phi_0\ket{0}$.

\hspace*{5mm} 
There is another important merit in this approach.\cite{Pauli-Brodsky}\ 
If we confine ourselves to the particle modes, then the positivity and 
  discreteness of the longitudinal momenta lead to a finite-dimensional 
  Hamiltonian:\footnote{Strictly speaking, this holds
  only in 1+1 dimensions. We do not have any restriction to transverse 
  directions. Nevertheless, the situation is much simpler than that 
  in the ET quantization. }\ 
The number of partitions of the total momenta $P^+$ into positive momenta 
  $P^+=\sum_n p^+_n$ is now finite. 
Thus, it is easy to diagonalize the Hamiltonian numerically, which gives 
  the eigenvalues and wavefunctions of excited states simultaneously. 

\hspace*{5mm} 
All the calculations are done with finite $L$ after that we take 
  the infinite volume  limit $L\to \infty$.

\vspace{-2mm}

\item{2.} {\it Setting infrared cutoffs for the longitudinal momenta}\\
The secondary method consists of eliminating the zero mode from the theory.
As we suggested above, if we set an infrared cutoff for the longitudinal 
  momentum as $p^+>\epsilon$, then we have no zero mode, and the vacuum 
  becomes trivial. 
In this case, the information concerning vacuum physics must be carried by 
  counterterms, which remove any infrared divergence.
If this is the case, we must perform ``nonperturbative'' 
  renormalization in the Hamiltonian formalism.
This is actually a very difficult task, due to the lack of the Lorentz 
  covariance.
It leads to a disastrous situation, with infinitely many 
  counterterms.\cite{Wilson} \  
To successfully treat them is a  challenging problem, and 
  some ideas (such as coupling coherence in the similarity 
  renormalization group method) have been developed, but we do not 
  discuss these in the present paper.
This is simply because one of our aims is to describe D$\chi$SB in the 
  Nambu--Jona-Lasinio (NJL) model with the non-renormalizable property.
As long as we consider the NJL model, we do not have to regard D$\chi$SB 
  as a problem of renormalization, though a regularization scheme is 
  a great matter in the NJL model.

\vspace{-2mm}

\item{3.} {\it Working off the light cone}\\
In order to regularize the zero-mode singularity, we can work slightly 
  off the LF and define the LF theory as a limit of the theory so 
  obtained.\cite{Near-LF} \ 
This can be achieved by choosing near-light-front coordinates 
  like $x^+=x^0\sin \theta  + x^3\cos \theta $ and 
  $x^-=x^0\cos \theta -x^3\sin\theta $ with $\theta\sim \pi/4$.
However, this method lacks the primary merit of the LF formalism, the 
  simplicity of the vacuum, and thus the amount of calculations it requires 
  is almost the same as, or greater than, that for the usual ET 
  calculation.
This approach is pedagogical and suggestive, but it is not 
  a LF theory, and simply for this reason, we do not consider it here 
  even though we learn much from it.\\

\end{description}

\vspace{-2mm}

\noindent We have been attacking the problem of describing vacuum 
  physics on the LF from the viewpoint of zero-mode 
  physics.\cite{Ita-Mae,Thesis,Itakura,I,II,Maedan,AH,Topology} \ 
In the present paper, based on our previous work, we focus on 
  the chiral symmetry breaking\cite{Ita-Mae,Thesis,Itakura,I,II} 
  and the effects of coupling to the gauge field.\cite{AH,Topology} \ 
We follow the DLCQ method when we have scalar fields.
Considerable efforts have been made with this method\cite{SSB}\  for 
  the spontaneous (discrete) symmetry breaking in a simple scalar model
 ($\lambda \phi_{1+1}^4$).
In these works, results have been obtained that are consistent with 
  the conventional results for the critical coupling as well as the 
  nonzero vacuum expectation value $\bra{0}\phi\ket{0}\neq 0$.
We see below that this method is also applicable to models with 
  chiral symmetry.
When we have only fermionic fields, we do not use
  any of the three methods described above.
However, we carefully consider the infrared singularities associated 
  with the longitudinal zero mode of the fermion bilinear.
In this sense, our approach in this case is physically similar 
  to the DLCQ method.

Finally, let us comment on another important issue.
When we discuss the chiral symmetry on the LF, we must be careful about 
  its special properties.
On the LF, the chiral transformation is defined differently 
  from the ordinary one.
The reason for this is the following.
Half of the degrees of freedom of any LF fermion are dependent variables
  to be represented by other independent variables.
Thus, it is natural to define the chiral transformation 
   only for the independent component of the fermion.
As we will discuss, it is not evident whether the LF chiral 
  transformation thus defined is equivalent to the ordinary one.
In fact, most surprisingly, it is not equivalent in the simplest case,
  as a massive free fermion theory is invariant under the LF chiral 
  transformation.\cite{Mustaki} \  
We discuss this problem in great detail.

With all these points in mind, we can rephrase the above questions in more 
   concrete terms as follows: \\

\vspace{-2mm}

\begin{description}
\item{(i)} {\it What is the difference between the LF chiral transformation 
      and the ordinary one? }   
      Even though we are ultimately interested in the breaking of the usual
      chiral symmetry, we must know the differences and similarities of the 
      two. This should be done both at classical and quantum levels.

\vspace{-2mm}

\item{(ii)} {\it How does a gap equation for the chiral condensate emerge? } 
      Quite generally, we need to resort to a gap equation in finding a 
      nonzero value of the order parameter. This should be the case
      even in the LF formalism. 

\vspace{-2mm}

\item{(iii)} {\it What is the consequence of the coexistence of a nonzero 
        chiral condensate and the trivial Fock vacuum? }
      We must understand why it is possible for the Fock vacuum to 
      give a nonzero condensate and how the theory should be modified.\\
      
\end{description}

\vspace{-2mm}

\noindent We give clear answers to these questions 
  based on the study of several models.
First,  in the next section, we highlight the above-mentioned 
  unusual aspects of the LF chiral transformation.
Using the free field theory as a simple example, we demonstrate the 
  difference between the LF chiral transformation and the usual one.
This partially answers Question (i).
More realistic consideration with interacting field theories is 
  given in Sect.~III.
There, we discuss the sigma model,\cite{Tsujimaru-Yamawaki}\ 
  the chiral Yukawa model,\cite{I} \ 
  and the NJL model\cite{II} \  as examples of a scalar theory, a scalar 
  and fermionic theory, and a purely fermionic theory, respectively.
We also investigate in each model how the gap equations for 
  the chiral condensate appear.
This corresponds to the answer to Question (ii).
Such a model-dependent study is inevitable to understand the chiral 
  symmetry breaking on the LF.
Then, we discuss in Sect.~IV common
  features in the methods of realizing D$\chi$SB on the LF.
It turns out that the chiral transformation of the dependent variables 
  are necessarily modified in the symmetry breaking phase.
Such modification significantly affects the system and resolves 
  the somewhat paradoxical situations regarding the trivial vacuum and chiral 
  symmetry breaking, and between the nonzero chiral condensate and the
  invariance of 
  the vacuum under the LF chiral transformation. 
This analysis should provide the answer to Question (iii).
In Sect.~V, we consider the effects of coupling to the gauge field.
As an example of gauge theories, we discuss the $U(1)$-gauged sigma 
  model, which is equivalent to the Abelian Higgs model.\cite{AH,Topology} \ 
A summary and conclusions are given in the last section. \\


\section{Chiral transformation on the light front}

In this section, we describe the basic ingredients of chiral symmetry 
  on the LF.  
Starting from the definition of the LF chiral transformation,
   we demonstrate how it is different from the ordinary one
   by using the free fermion theory.
A most remarkable difference is the fact that the massive free fermion 
  theory is invariant  under the LF chiral transformation.
(However, it turns out that this unusual property is highly exceptional.
  Indeed, it cannot be seen in any models we discuss.)
We also comment on the chiral transformation of the scalar field and 
  on the order parameter.\\

\subsection{Definition}

As we have already commented, the definition of the chiral transformation 
  on the LF is different from the usual one.
This is due to the special nature of the canonical structure of the LF 
  fermionic field.
To see this, it is sufficient to consider the kinetic term of the 
  Lagrangian ${\cal L}_0=\bar \Psi i\del\!\!\!/\Psi.$
Splitting the fermion field into the ``good'' and ``bad'' components as
\BQ
\Psi=\psi_++\psi_-,\quad \psi_\pm=\Lambda_\pm\Psi
\EQ
by using the projectors $\Lambda_\pm=\gamma^\mp\gamma^\pm/2$, 
  we easily find that the ``bad'' component $\psi_-$ is a dependent 
  variable: ${\cal L}_0=\sqrt2 \psi_+^\dag i\del_+ \psi_+ + 
  \sqrt2 \psi_-^\dag i\del_- \psi_-+\cdots$.
Note that $\del_-$ is a spatial derivative.
The canonically independent variable consists of only the ``good'' 
  component $\psi_+$.
It is convenient for practical calculations to use the ``two-component
  representation'' of the $\gamma$ matrices, so that the projectors 
  $\Lambda_\pm$ are expressed as
\BQ
\Lambda_+=\pmatrix{
{\bf 1} & 0 \cr
0 & 0 \cr
},\quad
\Lambda_-=\pmatrix{
0 & 0 \cr
0 & {\bf 1} \cr
}.\label{two-compo}
\EQ
Then, the projected fermions have only upper or lower components:
\BQ
\psi_+=\! \Lambda_+\Psi\equiv 2^{-\frac{1}{4}}\left (\matrix{
\psi \cr
0\cr
}\right ),\quad 
\psi_-=\! \Lambda_-\Psi\equiv 2^{-\frac{1}{4}}
\left (\matrix{
0\cr
\chi \cr
}\right ),\label{spinors}
\EQ
where we have defined the two-component spinors $\psi$ and $\chi$.
Among various representations satisfying Eq.~$(\ref{two-compo})$, 
  we use the following one in the present paper:
\BQ
\gamma^+=\sqrt2 \pmatrix{
0 & 0\cr
{\bf 1} & 0\cr
},\quad 
\gamma^-=\sqrt2
\pmatrix{
0 & {\bf 1} \cr
0 & 0 \cr
},\quad
\gamma^i=
\pmatrix{
-i\sigma^i & 0 \cr
0 & i\sigma^i \cr
},\label{two-compo-gamma}
\EQ
for $i=1,2$ and 
\BQ
\gamma_5=\pmatrix{
\sigma^3 & 0 \cr
0 & -\sigma^3 \cr
}.\label{gammafive}
\EQ
Now let us define the chiral transformation on the LF.
Since the bad component $\chi$ is a constrained variable in the LF formalism,
  it is subject to a constraint equation, what we call the 
  ``fermionic constraint'':
$$
i\del_-\chi =\frac{1}{\sqrt2}\left(-\sigma^i\del_i + m\right)\psi 
-\frac{\delta}{\delta \chi^\dag}{\cal L}_{\rm int}[\psi, \chi, \cdots].
$$
This is in general a nonlinear equation with respect to $\chi$.
Therefore, we cannot impose the chiral transformation freely on the 
  whole field.
We must impose it so that it is consistent with the constraint relation, 
  or apply it only on the good component from the beginning.
Since a consistency check is an involved task in interacting theories, 
  we just define the chiral transformation only for the good component:
\BQ
\psi\longrightarrow {\rm e}^{i\theta \sigma^3}\psi~. \label{Chiral_Transf}
\EQ
This is the definition of the LF chiral transformation.
If we do not use the specific two-component representation, 
 it is written $\psi_+\longrightarrow e^{i\theta \gfive }\psi_+$.
The transformation of the bad component can be found if we solve the 
  constraint equation.
We indeed do check this using some models in the next section.
However, before doing that, it is a good exercise to consider the 
  simplest case, that of a free fermion. \\

\subsection{Massive free fermion}

The most impressive and distinguishing property of the LF chiral 
  transformation is that the free fermion theory is invariant under the 
  transformation (\ref{Chiral_Transf}) {\it even if it has a mass term}.
Let us confirm this amazing fact explicitly at the Lagrangian level.
First of all, it is convenient to separate the solution of the fermionic 
  constraint $\chi=(\sqrt{2}i\del_-)^{-1}\left(-\sigma^i\del_i+m\right)\psi$ 
  into mass-independent and mass-dependent parts $\chi=\chi^{(0)} +\chi^{(m)}$ 
  as
$$
\chi^{(0)}=-\frac{1}{\sqrt2}\sigma^i \del_i \frac{1}{i\del_-}\psi~,
\qquad \chi^{(m)}=\frac{m}{\sqrt2}\frac{1}{i\del_-}\psi~,
$$
where $(i\del_-)^{-1}$ is defined in Appendix A.
The LF chiral transformation (\ref{Chiral_Transf}) induces
\BQ
\chi^{(0)}\longrightarrow {\rm e}^{-i\theta\sigma^3}\chi^{(0)}~,\quad
\chi^{(m)}\longrightarrow {\rm e}^{i\theta\sigma^3}\chi^{(m)} ~.\label{transfm}
\EQ
When $m=0$, we have $\chi^{(m)}=0$. 
Thus the chiral transformation of the full field $\Psi$ turns out to be 
  equivalent to the usual one [recall the representation of $\gamma_5$, 
  Eq.~(\ref{gammafive})]: 
\BQ
\Psi=2^{-1/4}\left(
\matrix{\psi\cr \chi^{(0)}}
\right),\quad 
\Psi\longrightarrow
{\rm e}^{i\theta\gamma_5}\Psi. \label{chi_transf_full}
\EQ
In this case, everything is the same as in the usual chiral transformation.

The situation is different for the massive case.
Here, the free fermion Lagrangian is compactly expressed as
${\cal L}_{\rm free}= \bar\Psi(i\del\!\!\!/-m)\Psi= 
          \psi^\dagger \omega_{\rm EOM} +\chi^\dagger\omega_{\rm FC},$ 
where 
  $\omega_{\rm EOM}=i\del_+\psi 
   -\frac{1}{\sqrt2}(\sigma^i\del_i+m)\chi=0$ 
  is the equation of motion for $\psi$ and 
  $\omega_{\rm FC}=i\del_-\chi -\frac{1}{\sqrt2}(-\sigma^i\del_i+m)\psi=0$ 
  is the fermionic constraint.
The second term here is zero, and it is invariant under the LF chiral 
  transformation.
Now, substituting $\chi=\chi^{(0)}+\chi^{(m)}$ into the Lagrangian, 
  we find that the first term $\psi^\dag \omega_{\rm EOM}$ is decomposed 
  into obviously invariant and (seemingly) non-invariant terms
$$
\psi^\dagger \omega_{\rm EOM}=\psi^\dagger  \left[i\del_+\psi 
-\frac{1}{\sqrt2}\left(\sigma^i\del_i \chi^{(0)}+m \chi^{(m)}\right) \right]
+\psi^\dagger \left[-\frac{1}{\sqrt2}
              \left(\sigma^i\del_i \chi^{(m)} +m \chi^{(0)}\right) 
              \right]~.
$$
The first term of this contribution consists of an $m$-independent term 
  and a  term  quadratic in $m$ while the second 
  term  depends linearly on $m$.
The apparent invariance of the ${\cal O}(m^2)$ term is very intriguing.
The ${\cal O}(m)$ term changes under the chiral transformation, 
  but thanks to a relation between $\chi^{(0)}$ and $\chi^{(m)}$,
\BQ
\sigma^i \del_i \chi^{(m)}+m\chi^{(0)}=0~,\label{relation}
\EQ
it eventually vanishes, and therefore the Lagrangian is totally invariant 
  even if we have a mass term.
Note that the invariance is closed within each term of ${\cal O}(m^n)$,
 and there is no mixing or cancellation between different order contributions.
Now that we have a symmetry, a corresponding conserved Noether current should 
  follow:\cite{Mustaki}
\BQ
j_5^{\mu}=\bar \Psi \Gmu \gfive \Psi 
   - m \bar \Psi \Gmu\gfive \frac{1}{i\del_-}\gamma^+\psi_+~.
\EQ
This, of course, reduces to the usual current in the massless limit, 
\BQ
J_5^\mu= \bar \Psi \Gmu \gfive \Psi. \label{Chiral_current}
\EQ
The chiral charge is given by
\BQ
Q^{\rm LF}_5=\int_{-\infty}^{\infty} dx^-d^2x_\perp 
  j_5^+ (x)
= \int_{-\infty}^{\infty} dx^-d^2x_\perp \psi^\dagger \sigma^3 \psi~.
\label{charge}
\EQ
Since $j_5^+=J_5^+$, the chiral charge is the same as that in
  the massless case.
This is reasonable, because the chiral transformation is defined 
  irrespective of whether we have a mass term.

In a quantum theory, the normal ordered chiral charge is given by
\BQ
:Q_5^{\rm LF}:=\int_0^\infty dk^+ \int_{-\infty}^\infty d^2k_\perp
  \left( b^\dag(\k)\sigma_3 b(\k) + d^\dag(\k)\sigma_3 d(\k)\right),
\EQ
where the mode expansion of the field is done at $x^+=0$, as in 
  Ref.~\citen{Leutwyler_meson}:
\BQ
\psi(x)=
\int_{-\infty}^{\infty}\frac{d^2k_\perp}{2\pi}
\int_0^{\infty}\frac{dk^+}{\sqrt{2\pi}}
\left[ ~
b(\k){\rm e}^{i{\scriptsize \k\x}}+d^{\dagger}(\k){\rm e}^{-i{\scriptsize \k\x}}
~\right],\label{mode_exp}
\EQ
with $\k\x\equiv -k^+x^-+k_\perp^ix_\perp^i$.
Note that the quantization has been performed only for the good component 
  $\psi_\alpha$ ($\alpha=1,2$ is a spinor index):
\BQ
\left\{\psi_\alpha(x),\ \psi_\beta^{\dagger}(y)\right\}_{x^+=y^+}
=\delta_{\alpha\beta}\delta(x^--y^-)\delta^{(2)}(x_\perp-y_\perp)~.
  \label{Quantization}
\EQ
If we use the mode expansion with helicity basis, $Q^{\rm LF}_5$ 
  measures the net helicity of a state.
Since the chiral charge does not contain a $b^\dag d^\dag$ term, it 
  annihilates the vacuum $\ket{0}$ irrespective of the existence of 
  a mass term:
\BQ
:Q_5^{\rm LF}:\ket{0}=0.
\EQ
This is a general property of the charge defined on the LF
  (null-plane charge).
Every charge operator is defined as the spatial integration of the 
  +-component of a current operator, which implies that the charge 
  operator has longitudinally zero momentum.
Thus the null-plane charge cannot contain any terms like
  $b^\dag d^\dag$ or $bd$ that cannot form zero longitudinal 
  momentum.\footnote{The fermionic field does not have zero modes because 
  we normally use antisymmetric boundary conditions in the 
  longitudinal direction.}\ 
Therefore, this charge {\it always} annihilates the 
  vacuum, irrespective of symmetry.\cite{Leutwyler} \ 
The same argument holds for the Hamiltonian, which is the charge of the 
  energy-momentum tensor. 
Thus, $:P^-:\ket{0}=0$ as long as we can ignore the zero modes.

The fact that the chiral current $j_5^\mu$ is a conserved current even for 
  $m\neq 0$ is a distinguishable property of the ``LF'' chiral 
  transformation.
However, if we are ultimately interested in the usual chiral symmetry 
  and its breaking, it would be better to treat $J_5^\mu$ instead of 
  $j_5^\mu$.
Here $J_5^\mu$ was defined in Eq.~(\ref{Chiral_current}) as a chiral 
  current for a massless fermion. 
If we use it even for the massive case, the divergence of $J_5^\mu$ 
  gives the ordinary relation
\BQ
\del_\mu J^\mu_5=2m \bar\Psi i\gamma_5 \Psi~. \label{current_div}
\EQ
This can be easily verified by using the Euler-Lagrange equation 
  for a massive fermion.
The unexpected symmetry of the massive fermion is certainly a unique and 
  intriguing property of the LF chiral transformation,
  but it is not clear whether such an unusual situation holds even for
  general interacting theories. 
If it is not the case in interacting models, we do not have to consider it.
In fact, as we will see in the next section, neither the chiral Yukawa model 
  nor the NJL model possesses this symmetry.\\

\subsection{Scalar field}

Let us comment on the chiral transformation of the scalar fields 
  ($\sigma, \pi$).
If we use the DLCQ method, we find a  situation similar to that for
  the chiral transformation of fermions.
Suppose that the system is invariant under the $U(1)$ chiral symmetry: 
\BQ
\left( \matrix{\sigma  \cr \pi} \right) \rightarrow   
\left(
       \matrix{ \cos 2 \theta   &  \sin 2 \theta  \cr
             -\sin 2 \theta   &  \cos 2 \theta        }
\right)
\left( \matrix{\sigma  \cr \pi} \right)~. \label{chi_transf_full_scalar}
\EQ
The same discussion as that given for the fermionic fields can be 
  applied to the case of scalar fields in the DLCQ approach.
This is because the zero modes are constrained variables to be determined 
  by nonzero modes (see Sect.~IIIA).
Hence, it is natural to define the chiral transformation 
  only for the oscillating modes: 
\BQ
\left( \matrix{\varphi_\sigma  \cr \varphi_\pi} \right) \rightarrow   
\left(
       \matrix{ \cos 2 \theta   &  \sin 2 \theta  \cr
             -\sin 2 \theta   &  \cos 2 \theta        }
\right)
\left( \matrix{\varphi_\sigma  \cr \varphi_\pi} \right)~,
\EQ
where $\sigma(x)=\sigma_0(x_\perp)+\varphi_\sigma(x)$ and 
      $\pi(x)=\pi_0(x_\perp)+\varphi_\pi(x)$.
The transformation of the zero modes should follow from the solutions to 
  the constraint relations for the zero modes.
More detailed discussion on this point is given in the next section.\\

\subsection{Order parameters}

Usually broken phase is characterized by nonzero 
  vacuum expectation values (VEVs) of the order parameters.
For example, the order parameters are the scalar field itself
  $\langle \sigma\rangle$ for the $U(1)$ symmetry in the sigma model 
  and the fermion bilinear operator $\langle\bar\Psi\Psi\rangle$ 
  for the chiral symmetry breaking.
In the usual ET formulation, such nonzero values are realized 
  by selecting a nontrivial vacuum that is different from the Fock vacuum. 
On the other hand, the vacuum in the LF formulation is determined 
  kinematically to be the Fock vacuum. 
This leads us to ask: How is it possible for those order parameters 
  to take nonzero values in the LF formalism?
A hint for the answer to this question lies in the fact that 
 {\it both order parameters contain  dependent variables in the 
  LF formalism}.
In the above two examples, they are 
 $\sigma=\sigma_0+\varphi_\sigma$ and 
\BQ
\bar\Psi \Psi =\frac{1}{\sqrt2}(\psi^\dag \chi + \chi^\dag \psi) ,
\EQ
where the zero mode $\sigma_0$ and the bad component $\chi$ are dependent 
  variables to be determined by the independent variables through 
  constraint relations.
Since the constraint equations reflect an interaction, 
  information concerning the dynamics enters the order 
  parameters in terms of these dependent variables.
This might lead to nonzero VEVs of the order parameters.  
We see that this is indeed the case. 
Note that, without information regarding the interaction, 
  order parameters are not able to have nonzero values.
In the ET formalism, this information is of course supplied by the new vacuum.
Since the LF vacuum is determined kinematically, it is natural to expect that
  the dynamical information is supplied by the solution of the constraint.
This is the only point with regard to which the information about the 
  interaction can enter the order parameter.

We must be careful about a use of the term ``order parameters'' 
  in the LF formulation.
Its meaning is a little different from the usual one.
To see this, recall the reason that $\langle \bar\Psi\Psi\rangle$ can be 
  considered as the order parameter in the ET formulation.
The reason is clear: 
The chiral transformation of the fermion bilinears is
\BQ
\left[ Q_5^{\rm ET}, \bar\Psi i\gamma_5\Psi \right]=-2i\bar\Psi\Psi 
\label{ChiralET}
\EQ
and therefore a nonzero value of $\langle \bar\Psi\Psi\rangle$ 
  immediately implies the violation of chiral symmetry.
On the other hand, as we saw above, any charge operator  in the LF formulation
  annihilates the vacuum, irrespective of symmetry.
Therefore if Eq.~(\ref{ChiralET}) held in the LF formulation,  
  a nonzero value of $\langle \bar\Psi\Psi\rangle$ would seem to be 
  prohibited on the LF.
Hence it is not clear whether the quantity $\langle\bar\Psi\Psi\rangle$ 
  plays the role of the order parameter in exactly the same way as 
  in the ET quantization.
In what follows, for practical purposes, we identify the phase according 
  to the value of  $\langle\bar\Psi\Psi\rangle$, because our eventual 
  interest is in the breaking of the ordinary chiral symmetry.
Hence, if we find $\langle\bar\Psi\Psi\rangle\neq 0$ in the LF formalism,
  we say that the system is in the broken phase. 
However, we must not forget about the above paradoxical situation.
We resolve this problem in a later section.\\


\section{Appearance of the gap equations: Model analysis}

The main aim of this section is to demonstrate how to obtain the gap equations 
  by using concrete models.
Recall that to find nontrivial gap equations is an indispensable step
  towards the symmetry breaking.
This is of course true of the LF formalism.
Here we see the unique way that they appear in the LF formalism.
The models we consider here are the sigma model, the chiral Yukawa model, 
  and the NJL model, which are examples of 
  a scalar theory,  a coupled system of scalar and 
  fermion fields,  and a purely fermionic theory, respectively.
What is common among these models is the appearance of the gap equations from 
  characteristic constraint equations of the LF formalism.
Various physical consequences, which follow after we obtain nonzero 
  condensates, are discussed in the next section.\\

\subsection{The sigma model -- Scalar fields}

The first example we consider is the linear sigma 
  model.\cite{Tsujimaru-Yamawaki} \  
Its simplest version, with only scalar fields and a potential with 
   the wrong sign for the mass term, is a basic and classic model of 
   spontaneous symmetry breaking.
Actually this ``tree-level'' symmetry breaking is not of interest presently.
We are interested in {\it dynamical} chiral symmetry breaking.
However, we can learn much from the analysis about how we should treat 
  the zero modes. 
The model is defined by
\BQ
{\cal L}=\frac12 \left\{(\del \sigma)^2+(\del \pi)^2\right\} 
         +\frac12 \mu^2 (\sigma^2+\pi^2)-\frac{\lambda}{4}(\sigma^2+\pi^2)^2 
         +c \sigma,
\EQ
where the last term is the explicit breaking term, and the potential has 
  the wrong sign for the mass term.
In order to treat the longitudinal zero mode appropriately, wek use
  the DLCQ method, in which we compactify the longitudinal space into a circle 
  with periodic boundary conditions on the scalars.
Then the zero modes are given as
\BQ
\sigma_0(x_\perp)=\frac{1}{2L}\int_{-L}^L dx^- \sigma(x), \quad
\pi_0(x_\perp)=\frac{1}{2L}\int_{-L}^L dx^- \pi(x), 
\EQ
which leads to a clear separation of the fields 
  $\sigma(x)=\sigma_0(x_\perp)+\varphi_\sigma(x)$ and 
  $\pi(x)=\pi_0(x_\perp)+\varphi_\pi(x)$.

As emphasized above, the most distinguishing feature of a scalar field 
  in the DLCQ method is that the zero mode is not an independent 
  degree of freedom.
This is because the momentum conjugate to the longitudinal zero mode 
  drops out of the Lagrangian.
(Note that the Lagrangian is linear in terms of the LF time derivative:
  $\frac12 (\del \sigma)^2= \del_+\sigma\del_-\sigma - 
  \frac12(\del_\perp\sigma)^2$.)
Thus, the zero modes are subject to the constraint relations
\BQA
&&\frac{1}{2L}\int_{-L}^L dx^- \left\{-(\mu^2+\del_\perp^2)\pi
               +\lambda \pi (\sigma^2+\pi^2)\right\}=0,\\
&&\frac{1}{2L}\int_{-L}^L dx^- \left\{-(\mu^2+\del_\perp^2)\sigma
               +\lambda \sigma (\sigma^2+\pi^2) \right\}=c.
\EQA
These are easily obtained from the longitudinal integration of the
   Euler-Lagrange equation, due to the property 
   $\int_{-L}^Ldx^- \del_-\del_+\phi=0$.
These relations  are called the ``zero-mode constraints''.
Since they are due to the structure of the kinetic term, 
  these constraints exist in any scalar models.
When we have only scalar fields, we always have the zero-mode 
  constraints.\footnote{Note that 
  this does not necessarily hold if gauge fields couple to scalars, 
  because the structure of the kinetic term changes. 
  We discuss this in Sect.~V. }\ 
They are of special importance for the description of the symmetry breaking.
To see this, let us decompose the longitudinal zero modes into 
  c-number parts and (normal-ordered) operator parts,
\begin{eqnarray}
&&\sigma_0=\sigma_0^{({\rm c})}
          +\sigma_0^{({\rm op})}(\varphi_\sigma,\varphi_\pi) ,
  \label{cae}\\
 && \pi_0 = \pi_0^{({\rm c})} 
          + \pi_0^{({\rm op})}(\varphi_\sigma,\varphi_\pi).
  \label{caeb}
\end{eqnarray}
If the c-number parts of the solutions are nonvanishing, 
   it directly follows that there are nonzero condensates: 
   $\langle 0 | \sigma | 0\rangle =\langle 0 | \sigma_0 | 0\rangle 
   =\sigma_0^{({\rm c})}\neq 0$ and 
   $\langle 0 | \pi | 0\rangle =\langle 0 | \pi_0 | 0\rangle 
   =\pi_0^{({\rm c})}\neq 0 $.
Therefore {\it it is necessary to find a nontrivial solution of the
   zero-mode constraint for describing the symmetry breaking}.
Here, it is important to recall that only zero modes can give
  nonzero values to the scalar fields (cf. Sect.~IID).

However, it is generally very difficult to solve the zero-mode constraint in 
   quantum theory.
The zero-mode constraints are nonlinear equations 
  among operators, and thus we must face the notorious problem 
  of operator ordering. 
Since the solution necessarily depends on the operator ordering, we must 
  specify an appropriate ordering with some criterion.
In many works involving the zero-mode constraints,
  people often choose, on general grounds, the Weyl ordering with 
  respect to both constrained and unconstrained variables.
However, it is not obvious whether the Weyl ordering in constraint
  equations makes sense.
The dependent variables are written in terms of the independent variables,
  and 
  thus if we substitute the solution into the constraint, 
  the final ordering  of the each term is no longer the Weyl ordering 
  with respect to the independent variables.
Instead, the most reliable criterion for determining the operator ordering
  is as follows.
To make the discussion clear, let us consider a commutator 
  $[\sigma_0, \varphi_\sigma]$.
This can be evaluated in two different ways,
(I) by using the solution $\sigma_0^{\rm sol}=
    \sigma_0(\varphi_\sigma,\varphi_\pi)$ of the zero-mode
    constraint with the standard quantization rule ($i,j = \sigma, \pi$)
\BQ
\left[ \varphi_i(x), \varphi_j(y)\right]|_{x^+=y^+}
=-\frac{i}{4}\left\{\epsilon(x^--y^-)-\frac{x^--y^-}{L}\right\}\delta_{ij}
\delta^{(2)}(x_\perp-y_\perp),
\EQ
or
(II) by calculating the Dirac bracket $\{\sigma_0, \varphi_\sigma\}_{\rm D}$
     and transforming it into the quantum commutator.
For case (I),  we assume that the solution was obtained without 
  using the commutator $[\sigma_0, \varphi_\sigma]$.
When we solve the constraint, we must work with a specific operator ordering,
  and the result depends on the ordering we choose.
For case (II), we must also determine the ordering on the right-hand side 
  of the Dirac bracket $\{\sigma_0, \varphi_\sigma\}_{\rm D}=\cdots$.
Thus, we have two ambiguities in the operator ordering: the ordering 
  of the constraint equation in (I) and the ordering of the right-hand 
  side of the Dirac bracket in (II).
Since the methods (I) and (II) must give the same result for 
  $[\sigma_0, \varphi_\sigma]$, the operator ordering  should be 
  imposed so that these two quantities are 
  identical.\footnote{If we cannot solve the 
  constraint without knowing the commutator $[\sigma_0, \varphi_\sigma]$, 
  then we must determine the operator ordering {\it self-consistently}.}
In other words, we determine the operator ordering of the right-hand side
  in the Dirac brackets so that it coincides with the direct evaluation.
This should be the criterion for an appropriate operator ordering.
However, as may be expected, it is extremely difficult to find such 
  a ``consistent operator ordering''.
Thus, practically, we just work with several particular
  orderings and compare the results to check the consistency.

In the sigma model with tree-level symmetry breaking, if we consider 
  only the leading order of the semiclassical approximation, we do not 
  have to worry about the operator ordering.
Then the zero modes are just classical numbers to be determined by 
\BQA
&&\mu^2\pi_0^{\rm cl} + \lambda (\pi_0^{\rm cl})^3 + \lambda \pi_0^{\rm cl} (\sigma_0^{\rm cl})^2=0,\label{sigma-const1}\\
&&\mu^2\sigma_0^{\rm cl} + \lambda (\sigma_0^{\rm cl})^3 + \lambda \sigma_0^{\rm cl} (\pi_0^{\rm cl})^2=c.\label{sigma-const2}
\EQA
When $c=0$, we have the usual symmetry breaking solution, 
  $\sigma_0^{\rm cl}=\sqrt{\mu^2/\lambda}$ and $\pi_0^{\rm cl}=0$.
If we solve the zero-mode constraints perturbatively with 
   respect to the coupling constant $\lambda$, 
   we do not have a symmetry breaking solution.
Because Eqs. (\ref{sigma-const1}) and (\ref{sigma-const2}) determine the 
   value of the condensate, we can say that they are the gap equations.
Indeed,  the same equations are also obtained by differentiating the 
   tree-level effective potential of spatially constant fields. 

The lesson we learn from the above analysis is that to realize the symmetry 
  breaking in the DLCQ approach, it is necessary to obtain a 
  nontrivial solution to the zero-mode constraint.
Since we have selected the potential so that it induces tree-level 
  symmetry breaking, even the classical solution of the zero-mode constraint 
  gives a nontrivial solution.
However, in general,  we have to solve the zero-mode constraint
  nonperturbatively at the quantum level.  \\


\subsection{The chiral Yukawa model -- Scalar and fermionic 
            fields}

Now let us study a more complicated example, the chiral Yukawa model.
We add an $N$-component fermion field $\Psi_a$ $(a=1,\cdots, N)$ that
  couples to the scalars via the Yukawa interaction:
\BQ
  {\cal L} = \bar \Psi_a ( i \rlap/ \partial - m )\Psi_a 
             + {N \over 2 \mu^2} \left\{( \partial\sigma)^2
                   + (\partial \pi)^2\right\}
             -{N \over 2 \lambda} (\sigma^2 +\pi^2)
                -(\sigma \bar \Psi_a \Psi_a 
                     + \pi \bar \Psi_a i \gamma_5 \Psi_a ).
  \label{aa}
\EQ
Here $\mu$ is a dimensionless parameter and the fermionic field is
  assumed to be antiperiodic in the $x^-$ direction.
This model is not a simple extension of the previous sigma model.
Note that the Lagrangian (\ref{aa}) does not have a quartic interaction 
  of the scalars.
Therefore, this model undergoes chiral symmetry breaking 
  at the quantum level.
This can be explicitly verified in a conventional calculation with
  the effective potential.
Also this may be seen clearly by the fact that the model becomes
   the NJL model in the infinitely heavy mass limit for scalars
   $ \mu \rightarrow \infty $ and that the NJL model exhibits D$\chi$SB 
   at the quantum level.\footnote{Furthermore, the chiral Yukawa model 
   in the mean field approximation is equivalent to the NJL model.}
The NJL model is discussed in the next subsection.

As we have scalar fields, let us use the DLCQ approach again.
There is one additional complication here that does not exist in
  the previous case.
In addition to the two zero-mode constraints for $\sigma_0$ and $\pi_0$,
\BQ
\left( \frac{\mu^2}{\lambda} - \partial_\bot^2 \right)
       \left(\matrix{\sigma_0 \cr \pi_0}\right)
     + \frac{\mu^2}{N} \frac{1}{\sqrt{2}}\int_{-L}^L \frac{dx^-}{2L}
           \left[ 
           \psi^{\dag}_a 
           \left(\matrix{1\cr -i\sigma_3}\right)
           \chi_a 
            + \chi^{\dag}_a 
           \left(\matrix{1\cr i\sigma_3}\right)
           \psi_a 
           \right]=0, 
  \label{acc}
\EQ
we have one fermionic constraint for the bad component $\chi$,  
\begin{equation}
   i \partial_- \chi_a(x) = {1 \over \sqrt2}\left( -\sigma^i\partial_i
             + m +\sigma(x) + i \pi(x) \sigma_3 \right) \psi_a ~.
  \label{acaa}
\end{equation}
These three equations form a coupled set for the dependent variables 
   $\sigma_0$, $\pi_0$ and $\chi$ to be represented by the other 
   independent variables, $\varphi_\sigma$,  $\varphi_\pi$ and $\psi$.

It is not difficult to obtain {\it classical} solutions to the above 
  equations.
To solve them classically, we treat the fermionic fields just as 
  Grassmann numbers.
The classical solutions enable us to check explicitly 
  whether the system with a {\it mass term} is invariant  or not 
  under the LF chiral transformation.
It turns out that chiral symmetry exists if and 
  only if $m=0$ and that the scalars and fermions transform as in 
  Eqs.~(\ref{chi_transf_full_scalar}) and (\ref{chi_transf_full}). 
Therefore, we do not have to consider the extra symmetry which is
  present in the massive free fermion theory.
Now that we can treat the LF chiral transformation as being equivalent 
  to the usual one, the chiral current is also given as usual.
Note that the LF chiral charge defined by spatial integration of its 
  plus component is written only in terms of independent degrees of  
  freedom (here, $\varphi_\sigma, \varphi_\pi$, and $\psi$), 
  which should be the case, by definition.
We do not discuss about the classical analysis any further here.
Explicit demonstrations involving the classical 
  solutions and their consequences are found in Ref.~\citen{I}. 
Below, we discuss similar situations in more detail 
  in the NJL model.


Now let us turn to the quantum theory.
Unlike the classical theory, it is extraordinarily difficult to solve 
  the constraints in this case as operator equations.
The primary reason for this is the problem of operator ordering.
Probably the best way to attack this problem is to find a consistent 
  operator ordering, as we discussed in the case of sigma model.
This is, however, a very difficult task in our model.
We, rather, consider only a specific ordering. 
It can be shown that the leading order (in the $1/N$ expansion) 
  result does not change even if we choose other operator orderings.

We must solve the coupled equations  (\ref{acc}) and (\ref{acaa}) 
  of the dependent variables $\chi$, $\sigma_0$ and $\pi_0$.
First of all, it is easy to remove $\chi$ from them. 
Solving the fermionic constraint (\ref{acaa}) formally by inverting $i\del_-$ 
  and then inserting the solution $\chi$ into Eq.~(\ref{acc}), 
  we find complicated 
  equations for the scalar zero modes $\sigma_0$ and $\pi_0$.
After changing the ordering, so that the leading order calculation becomes 
  relatively easy, we obtain
\begin{eqnarray}
  & &\hspace{-0.5cm}\Big({\mu^2 \over \lambda}-\partial_\perp^2 \Big)
      \left(\matrix{\sigma_0\cr \pi_0} \right)\nonumber\\
  &+& {\mu^2 \over 4N}
      \int_{-L}^{L}\!\! \frac{dx^- dy^-}{2L}
      \frac{\epsilon (x^- -y^-)}{2i}         
\left\{
        \psi^{a \dag}_x\left(\matrix{-1\cr i\sigma_3}\right)
        \sigma^i \partial_i \psi^a_y
     - \partial_i \psi^{a \dag}_y
        \left(\matrix{-1\cr i\sigma_3}\right)\sigma^i\psi^a_x
                      \right.                      \nonumber  \\
  & &+  \left(\matrix{m+\sigma(y)\cr \pi(y)} \right)
       \left( \psi^{a \dag}_x \psi^a_y
                       - \psi^{a \dag}_y \psi^a_x\right)
   -  \left. \left(\matrix{-\pi(y)\cr m+\sigma(y)}\right)
       \left( \psi^{a \dag}_x i\sigma_3\psi^a_y
       + \psi^{a \dag}_y i\sigma_3\psi^a_x
       \right)
       \right\} \nonumber\\
  &+ &       \ {\rm H.c.}=0,
  \label{caa}
\end{eqnarray}
where we use the definitions $\psi^a_y=\psi^a(y^-,x_\perp)$, 
  $\sigma(y)=\sigma_0(x_\perp)+\varphi_\sigma(y^-,x_\perp)$,  
  and so on.    
(Do not confuse the Pauli matrices $\sigma^i$ with the scalar field 
  $\sigma(x)$.)

We now show how to determine the leading order solution of the zero-mode 
  constraints in the $1/N$ expansion: 
  $\sigma_0=\sigma_0^{\rm (c)}+\sigma_0^{\rm (op)}={\cal O}(N^0)$ 
  and $\pi_0=\pi_0^{\rm (c)}+\pi_0^{\rm (op)}={\cal O}(N^0)$.
First of all, the c-number part of the zero modes can be chosen as 
  $\sigma_0^{({\rm c})} \neq 0$ and $\pi_0^{({\rm c})}=0$, 
  because we confirmed in the classical analysis that 
  $(\sigma_0, \pi_0)$ rotates chirally in the massless case.
Because of the relation which holds in leading order,
  $\langle \sigma \rangle = -\frac{\lambda}{N}\langle \bar \Psi \Psi \rangle$  
  and $\langle \pi \rangle 
  =-\frac{\lambda}{N}\langle \bar \Psi i\gamma_5 \Psi \rangle$, 
  a nonzero value of $\sigma_0^{\rm (c)}$ immediately implies a nonzero 
  fermion condensate.
Taking the VEV of the zero-mode constraint for $\sigma$, we find
\begin{equation}
  M-m
  = 2\lambda M \int \frac{d^2 p_\bot}{(2 \pi)^2} \sum_{n=\frac12,\frac32\cdots}
       { 1 \over p_n^+} \frac{\Delta p^+}{2 \pi}  ,
  \label{cai}
\end{equation}                     
where $\Delta p^+=\pi/L$ and  $p^+_n=\pi n/L$, and we have introduced 
  $M \equiv m +\sigma_0^{(\rm c)}$.  
Physically, this equation should be the gap equation,
    by which we can determine 
   the condensate $\sigma_0^{\rm (c)}$, and equivalently the physical 
   fermion mass $M$. 
However, it is not evident whether we can regard it as the gap equation,
   because Eq.~(\ref{cai}) in the naive chiral limit $m\rightarrow 0$ cannot 
   give nonzero $M$.
The flaw of this observation lies in the loss of mass dependence in the 
   infinite summation of Eq.~(\ref{cai}). 
It turns out\cite{Ita-Mae} that this can be fixed by supplying mass 
   information properly through a cutoff when we regularize the divergent 
   summation (see Sect.~IVA for more details).
For example, if we adopt a cutoff which respects the parity 
   invariance,\footnote{The parity transformation is given by 
    $p^+\leftrightarrow p^-,\ p_\perp\to -p_\perp$.}\  
   the infinite momentum sum can be approximated by an integral over $p^+$
   with the integration range $(M^2+{\p}_\perp^2)/2\Lambda <p^+<\Lambda$.   
Consequently,  Eq.~(\ref{cai}) becomes {\it dependent} on the mass $M$ and
   can be considered as a gap equation:
\begin{equation}
  M-m = \lambda M \frac{\Lambda^2}{4 \pi^2}
         \left\{ 2- \frac{M^2}{\Lambda^2} \left(
            1+{\rm ln}\frac{2 \Lambda^2}{M^2}
                \right) \right\}  .
  \label{caic}
\end{equation}
Indeed, even in the chiral limit $m \rightarrow 0$, this equation 
  is a nonlinear equation in $M$, and when
  the coupling constant $\lambda$ is larger than the 
  critical value $\lambda_{\rm cr}=2 \pi^2/\Lambda^2$, 
  we have a nontrivial solution $ M= M_0\neq 0$. 
This also implies that the c-number part of $\sigma_0$ has been determined 
  to be $\sigma_0^{({\rm c})} =M-m$.

When we derived the above cutoff, we used the dispersion relation with
  the dynamical fermion mass $M$. 
That is, we used $2p^+p^--{\p}_\perp^2=M^2$ instead of 
  $2p^+p^--{\p}_\perp^2=m^2$.
This is an important step in obtaining the gap equation, and it 
  corresponds to imposing the self-consistency.       
As we see below, the use of the physical fermion mass is natural, 
  since the prefactor of the integral in Eq.~(\ref{cai}) is $M$.

The operator parts of the zero modes are obtained by ``linearization'' of the
  zero-mode equations to leading order in $1/N$ expansion.
The result is 
\begin{equation}
\left(\matrix{\sigma^{(\rm op)}_0\cr\pi^{(\rm op)}_0}\right)
 = -{\mu^2 \over N} \left( m_{\rm ZM}^2 - \partial_\bot^2 \right)^{-1}
   \frac{1}{2L}\int_{-L}^Ldx^- \left[ : 
     \bar \Psi^a_{M}\left(\matrix{1\cr i\gamma_5}\right)
          \Psi^a_{M} : \right] ,
  \label{cea}
\end{equation}
where $m_{\rm ZM}^2 = \mu^2 m/(\lambda M)$ and $\Psi_M$ is a 
  free fermion operator with mass $M$:
\BQ
\Psi_M=2^{-\frac14}
\pmatrix{\psi\cr \chi_M}
=2^{-\frac14}
\pmatrix{\psi\cr \frac{1}{\sqrt2 i\del_-}(-\sigma^i\del_i + M )\psi}.
\label{massive_fermion}
\EQ
Finally, by inserting the c-number and operator parts of $\sigma_0$ and 
  $\pi_0$ into the solution for the fermionic constraint, we obtain the 
  bad component of the fermion:
\begin{equation}
\chi=\chi_M+\frac{1}{\sqrt2} \frac{1}{i\partial_-}
 \left(\sigma_0^{\rm (op)}-\pi_0^{\rm (op)}i\sigma_3 \right)\psi.
 \label{FC_sol_Yukawa}
\end{equation}                                                   
Equations (\ref{cea}) and (\ref{FC_sol_Yukawa}) and the c-number part 
  $\sigma_0^{({\rm c})}\neq 0,\ \pi_0^{({\rm c})}=0$ 
  are the leading-order solutions to the coupled equations.
If we select the nontrivial solution (i.e., a solution $M\neq 0$ even in the 
  chiral limit) of the gap equation, then the resulting theory should 
  describe the broken phase. 
The trivial solution gives the symmetric phase.
Therefore, the resulting theories written in terms of independent variables 
  change according to the solutions we choose.
Various consequences of this unusual situation are discussed in the 
  next section.

In summary, what we found in this analysis is that (A) the zero-mode 
  constraints are very important, since they contain information 
  about the gap equation, and that (B) we have to be careful about 
  the divergent summation over the longitudinal momenta to get a 
  nontrivial gap equation.
The second point is discussed in more detail in the next section.
The first point, that is, the metamorphosis of the zero-mode constraint into 
  the gap equation, has already been seen in the sigma model.
Here we find a situation similar to that in the more complicated system with
  chiral symmetry.
The appearance of the gap equation is more nontrivial in this model in 
  the sense that we need a quantum and nonperturbative analysis and a 
  prescription for the mass information loss.\\


\subsection{The Nambu--Jona-Lasinio model -- Fermionic 
            fields }

The last example in this section is the NJL model,\cite{NJL}\ 
  which is a fermion theory with four-Fermi interaction:
\BQ
{\cal L}=\bar\Psi(i\del\!\!\!/ -m)\Psi +
\frac{\lambda}{2N}\left[(\bar\Psi\Psi)^2+(\bar\Psi i \gamma_5 \Psi)^2\right]~.
\EQ
This model is the most well-known classic example exhibiting chiral 
  symmetry breaking.\cite{NJL_review}\ 
The chiral symmetry that is present in the massless Lagrangian dynamically 
  breaks down due to a nonzero condensate $\langle\bar\Psi\Psi\rangle \neq 0$.
Recently, there have been several efforts to reproduce this  in the 
  LF NJL model.\cite{Heinzl,Thesis,Itakura,II,Bentz}\ 
Here we explain one of such attempts based on our 
  studies.\cite{Thesis,Itakura,II}

Since we do not have scalar fields, the method based on the zero-mode 
  constraint is useless in this model.
However, it is very suggestive to consider the chiral Yukawa model from a 
  different point of view. 
Recall that the first task faced in the application of that model was 
  to remove the bad component 
  from the coupled constraint equations Eqs.~(\ref{acc}) and (\ref{acaa}).
This led us to the zero-mode constraints (\ref{caa}), and we successfully 
  extracted the gap equation from them.
However, we should note that we could have removed the zero modes first 
  (instead of the bad spinor component) from the coupled equations.
Indeed we can formally solve the zero-mode constraints (\ref{acc}) 
  and substitute the solutions into the fermionic constraint (\ref{acaa}).
Then {\it the resulting fermionic constraint must carry information 
  concerning the gap equation}, because the information should be preserved.
What does this tell us about the NJL model?
The LF NJL model has one complicated fermionic constraint that is 
  immediately obtained as the Euler-Lagrange equation for $\chi$,
\BQ
i\del_-\chi_a=\frac{1}{\sqrt{2}}\left(-\sigma^i\del_i+m\right)\psi_a
-\frac{\lambda}{2N}\left\{
   \psi_a\left(\psi^\dagger_b\chi_b+\chi^\dagger_b\psi_b\right)+
   \sigma^3\psi_a\left(\psi^\dagger_b\sigma^3\chi_b
                    -\chi^\dagger_b\sigma^3\psi_b\right)
\right\},\label{FC}
\EQ
where summation over the ``color'' and spinor indices is implied.
The above discussion for the chiral Yukawa model suggests that this 
  fermionic constraint contains the information about the gap equation. 
We will see that this is indeed the case.
In particular, this can be easily confirmed if we note the equivalence 
  of the NJL model and the heavy mass limit of the chiral Yukawa model.
Here, we do not use the DLCQ method, as we have no scalar fields.
The longitudinal extension is not compactified, but we impose the antiperiodic 
  boundary condition in this direction $\Psi(x^-=-\infty)=-\Psi(x^-=\infty)$.\\

\subsubsection{Classical solution of the fermionic constraint}  
Before proceeding to quantum analysis, let us consider the classical 
  solution of the fermionic constraint.
Since we can treat all the variables as Grassmann numbers in the classical
  analysis, the equation becomes tractable, and it is not  
  difficult to solve it. 
Indeed, the exact solution with antiperiodic boundary condition 
 is given by 
\BQ
\pmatrix{
\chi_{1a}(x)\cr
-\chi_{2a}^\dagger(x)
}=\frac{1}{\sqrt2}\int_{-\infty}^\infty dy^- G_{ab}(x^-,y^-,x_\perp)
\pmatrix{
m\psi_{1b}(y^-)-\del_z\psi_{2b}(y^-)\cr
-\del_z\psi_{1b}^\dagger(y^-)+m\psi_{2b}^\dagger(y^-)
},\label{classical_solution}
\EQ
where $\del_z=\del_1-i\del_2$, and 
the ``Green function'' $G_{ab}(x^-,y^-,x_\perp)$ is 
\BQA
&&G_{ab}(x^-,y^-,x_\perp)
=G^{(0)}(x)\left[\ \frac{1}{2i}\epsilon(x^--y^-)+C\ \right]G^{(0)}(y)^{-1}~,
\NN
&&G^{(0)}(x)={\rm P}\ {\rm e}^{i\frac{\lambda}{N}\int_{-\infty}^{x^-}{\cal A}(y^-)dy^-},
\quad {\cal A}_{ijab}=\pmatrix{
\psi_{1a}\psi_{1b}^\dagger & \psi_{1a}\psi_{2b} \cr
\psi_{2a}^\dagger\psi_{1b}^\dagger & \psi_{2a}^\dagger \psi_{2b} \cr
}~.\nonumber
\EQA
The integral constant $C$ is determined so that the solution satisfies 
  antiperiodic boundary condition.
The symbol ``P'' in the definition of $G^{(0)}(x)$ represents the 
  path-ordered product.

The most significant benefit of having the exact form of the classical 
  solution is that we can explicitly check the transformation 
  of the bad component 
  under the LF chiral transformation (\ref{Chiral_Transf}).
Recall the importance of the classical solution in the massive free fermion 
  theory in Sect.~IIB.
We apply similar analysis here.
Decomposition of $\chi$ is straightforward:
\BQA
\pmatrix{
 \chi^{(0)}_{1a}\cr
 -\chi^{(0)\dagger}_{2a}
 }
&=&-\frac{1}{\sqrt2}\int_{-\infty}^{\infty} dy^- G_{ab}(x^-,y^-,x_\perp)
\pmatrix{
 \del_z\psi_{2b}\cr
 \del_z\psi_{1b}^\dagger},\label{chi0}\\
\pmatrix{
\chi^{(m)}_{1a}\cr
-\chi^{(m)\dagger}_{2a}
}
&=&\frac{m}{\sqrt2}\int_{-\infty}^{\infty} dy^- G_{ab}(x^-,y^-,x_\perp)
\pmatrix{
\psi_{1b}\cr
\psi_{2b}^\dagger
}.\label{chim}
\EQA
Since the matrix ${\cal A}$, and thus $G_{ab}(x,y)$, is invariant under 
  the transformation (\ref{Chiral_Transf}), it is easy to find that 
  $\chi^{(0)}$ and $\chi^{(m)}$ transform as in Eq.~(\ref{transfm}).
Therefore, if $m=0$, the LF chiral transformation (\ref{Chiral_Transf}) 
  is equivalent to the usual chiral transformation.
The chiral current and the chiral charge are given by Eqs. 
  (\ref{Chiral_current}) and (\ref{charge}), respectively.

Hence the massless case is seen to be quite satisfactory. 
Then, what about the massive case?
As we discussed in Sect.~IIB, the chiral symmetry survives even with a mass 
  term in the free fermion case.
We must bear in mind such a possibility even in the NJL model. 
Thus it is worthwhile to check whether the {\it massive} 
   NJL model is invariant under the LF chiral transformation. 
To see this, it is convenient to treat the Hermite Lagrangian 
$$
{\cal L}_{\rm Hermite}=\frac{1}{2}i \psi^\dagger 
\buildrel\leftrightarrow\over{\del}_+\psi 
-\frac{1}{2\sqrt{2}} \left[
  \left( \psi^\dagger \sigma^i\del_i\chi
  +\del_i\chi^\dagger \sigma^i \psi   \right)
  +m\left(\psi^\dagger\chi + \chi^\dagger\psi \right) \right] ~.
$$
Note that this is equivalent to the free Lagrangian, except that 
  $\chi$ is a solution of Eq.~(\ref{FC}). 
Now, the apparently non-invariant term depends linearly on $m$:
$$
-\frac{1}{2\sqrt2}\psi^\dagger 
  \left(\sigma^i\del_i\chi^{(m)}+m\chi^{(0)}\right) 
+ {\rm H.c.}~.
$$
In the massive free fermion case, we had the same term, but 
  it eventually vanished due to Eq.~(\ref{relation}). 
However, in the NJL model, it is evident from Eqs.~(\ref{chi0}) and 
  (\ref{chim}) that such a relation does not hold, because 
  $G$ depends on $x_\perp$.
Therefore we have verified that the massive NJL model is {\it not} 
  invariant under the LF chiral transformation.
If and only if $m=0$, the LF chiral transformation is the symmetry 
  of the NJL model and is equivalent to the usual chiral transformation. 
This is, of course, not a surprising result, but it must be checked explicitly.

Irrespective of whether we have a mass term or not, we always use the 
  definition for the chiral current Eq.~(\ref{Chiral_current}).
In the massless case, it is, of course, a conserved current 
  $\del_\mu J^\mu_5=0$, while in the massive case, the usual relation 
$\del_\mu J^\mu_5=2m \bar\Psi i\gamma_5 \Psi$ holds,  
which is derived by using the Euler-Lagrange equation of the massive NJL model.
\\

\subsubsection{Quantum and nonperturbative solution}

Now let us turn to the quantum aspects of the model. 
We solve the nonlinear operator equation (\ref{FC}) by using 
  the $1/N$ expansion.
This entails two additional problems, as well as the usual operator 
  ordering problem. 
The first problem is determining how to count the order 
  ${\cal O}(N^n)$ of an operator instead of its matrix element.
The second is that it is physically difficult to justify the $1/N$ expansion 
  of the fermionic field itself.
To overcome these, in Ref.~\citen{II}, we rewrote the fermionic constraint 
  in terms of bilocal fields and solved them with fixed operator ordering.
More precisely, we introduced three bilocal fields defined by 
\BQA
&&\M_{\alpha\beta}(\x, \y)= 
\sum_{a=1}^N \psi^{a \dagger}_\alpha(x^+,\x)\psi^a_\beta(x^+,\y),\NN
&&\T_{\alpha\beta}(\x, \y) =
\frac{1}{\sqrt{2}}\sum_{a=1}^N \left( \psi^{a \dagger}_\alpha(x^+,\x)
\chi^a_\beta(x^+,\y)
 + \chi^{a \dagger}_\beta(x^+,\y) \psi^a_\alpha(x^+,\x) \right),\NN
&&\U_{\alpha\beta}(\x, \y) =
\frac{-i}{\sqrt{2}}\sum_{a=1}^N \left(
  \psi^{a \dagger}_\alpha(x^+,\x) \chi^a_\beta(x^+,\y) -
  \chi^{a \dagger}_\beta(x^+,\y)
\psi^a_\alpha(x^+,\x)
   \right).\nonumber
\EQA
The $1/N$ expansion of the bilocal operator $\M_{\alpha\beta}(\x, \y)$ is 
  known as the Holstein-Primakoff expansion, which is a special case of 
  the boson expansion method.\cite{BEM_itakura}\  
Finding $\T_{\alpha\beta}$ and $\U_{\alpha\beta}$ corresponds to 
  obtaining $\chi$ and $\chi^\dagger$. 
At each order of the $1/N$ expansion, the bilocal fermionic constraints are 
  linear with respect to $\T_{\alpha\beta}$ and 
  $\U_{\alpha\beta}$, and thus are easily solved.\cite{II}\  
 
In particular, the leading order of the bilocal fermionic constraint 
  reduces to the very simple form
\BQ
M-m=\lambda M\int \frac{d^3\p}{(2\pi)^3}\frac{\epsilon(p^+)}{p^+}~,
\label{lowest_const_NJL}
\EQ
where we have defined the physical fermion mass $M=m- (\lambda / N )\langle 
   \bar\Psi\Psi \rangle$, and $\langle\bar\Psi\Psi \rangle$ is the 
  leading order of $\T_{\alpha\alpha}(\x,\x)={\cal O}(N)$.
This is essentially the same as Eq.~(\ref{cai}), and therefore it
  becomes the gap equation (\ref{caic}) if we use the same cutoff scheme.
The solution of Eq.~(\ref{lowest_const_NJL}) gives the lowest order of 
  $\T_{\alpha\alpha}(\x,\x)$.
Similarly, higher order contributions are determined order by order.
Once we find solutions of the fermionic constraints, the Hamiltonian,
   which is also written in terms of the bilocal fields, is obtained. 
If we choose a nontrivial solution of the gap equation, 
  the resulting Hamiltonian describes the broken phase even 
  with a trivial vacuum.

In this last example, we have explained that the physical role of the 
  fermionic constraint is very similar to that of the zero-mode constraint 
  in the previous examples.
We have seen a close parallel between these two constraints.
In particular, it should be noted that the gap equation results from the 
  longitudinal zero mode of the bilocal fermionic constraint.
It is very natural that we can realize the broken phase by solving 
  the quantum fermionic constraint with the $1/N$ expansion,
  because the fermionic constraint is originally a part of the 
  Euler-Lagrange equation and thus must include relevant information 
  regarding the dynamics.
What we have done here is actually very similar to the usual mean-field 
  approximation for the Euler-Lagrange equations.
Indeed, the leading order in the $1/N$ expansion corresponds to the 
  mean-field approximation.
However, our method of solving the fermionic constraint with the boson 
  expansion method can easily go beyond the mean-field level.
Such a higher-order calculation enables us to derive the correct broken
  Hamiltonian, and so on. These points are discussed in Ref.~\citen{II} 
  in more detail.\\


\section{Chiral symmetry breaking on the light front}

In this section we discuss some unusual but important aspects 
  of the methods of realizing chiral symmetry breaking on the LF.
We found in the previous section that the chiral condensate 
  can be obtained from the characteristic constraint equations.
In spite of having a nonzero chiral condensate, our vacuum is still 
  the trivial Fock vacuum.
This coexistence of the chiral condensate and the trivial vacuum 
  leads to various unusual consequences.
After we make a more detailed investigation of the chiral condensate,
  and, in particular, the problem of the loss of mass information, 
  we introduce the concept of multiple Hamiltonians in the broken phase 
  and show how the Nambu-Goldstone bosons appear.
Then, we discuss the most important properties of the broken phase,
  that is, the peculiarity of the chiral transformation and the
  nonconservation of the chiral charge.\\

\subsection{Chiral condensate and problem of mass information loss}

Let us first consider the reasons why we succeeded in obtaining 
  a nonzero fermion condensate.
To this end, it is very instructive to see the fermionic constraint 
  (\ref{FC}) from a different point of view.
Rewriting the fermionic constraint as
$$
i\del_-\chi_a
 =\frac{1}{\sqrt{2}}\left(-\sigma^i\del_i+m \right)\psi_a
   -\frac{\lambda}{\sqrt2 N}
   \left( \psi_a \T_{\alpha\alpha}(\x,\x) 
  + i\sigma^3\psi_a \sigma^3_{\alpha\beta}\U_{\alpha\beta}(\x,\x) \right)~,
$$
and substituting the leading order solutions into the bilocal operators 
  $\T_{\alpha\alpha}(\x,\x)$ and $\U_{\alpha\beta}(\x,\x)$,
  we find that the leading order equation turns out to be equivalent to the
  fermionic constraint for a free fermion with mass $M$:
\BQ
i\del_-\chi_a =\frac{1}{\sqrt{2}}\left(-\sigma^i\del_i+M\right)\psi_a ~.
\EQ
Also at the same order, the equation of motion for the good component 
  $\psi$ tells us that the fermion acquires a mass $M$.
This procedure is actually equivalent to the mean-field approximation 
  in the Euler-Lagrange equation.
The same gap equation as before can be derived as a self-consistency 
  equation $M=m+(\lambda/N)\langle\bar\Psi\Psi\rangle$:\cite{Heinzl} 
\BQA
M&=&m+\frac{\lambda}{N}\cdot \frac{1}{\sqrt2}
\bra{0}(\psi^\dag\chi+\chi^\dag\psi)\ket{0}\NN
 &=&m+\frac{\lambda }{ N}\cdot \frac{M}{2}\bra{0}
 \left\{\psi^\dag\frac{1}{i\del_-}\psi 
-\left(\frac{1}{i\del_-}\psi^\dag \right)\psi\right\}\ket{0}\NN
 &=&m - \frac{\lambda}{N}\cdot  \frac{NM}{2} \int 
\frac{d^3\p}{(2\pi)^3}\frac{\epsilon(p^+)}{p^+}.
\label{mean_field}
\EQA
In the last line, we have followed the standard canonical LF quantization
  with the mode expansion (\ref{mode_exp}).
Two things are important here.
First, the right-hand side of the gap equations 
  (\ref{cai}) or (\ref{lowest_const_NJL}) comes essentially from the 
  vacuum expectation value of $\bar\Psi\Psi$ in a free massive theory.
Second, the fermion condensate $\langle \bar\Psi\Psi\rangle$ 
  is not zero for a massive fermion even if the vacuum is trivial.

Recall that Eq.~(\ref{mean_field}) cannot be the gap equation as it stands, 
  because it does not have the correct mass dependence.  
Therefore, we can now understand the essential question to be asked:
How can we compute $\bra{0}\bar\Psi\Psi\ket{0}$ correctly in the LF formalism?
This is not a problem limited to chiral symmetry breaking, but is more
  general.
Indeed, it is a common problem for the computation of more than 
  two-point functions. 
The case we encounter here is a special case of the fermion bilinears.
Let us consider this situation more rigorously.
In the massive free fermion theory, we can directly compute the two-point 
  functions of the fermion bilinears.
For example, the usual ET quantization scheme gives\cite{Schweber} 
\BQA
\bra{0} \bar\Psi (x) \Psi(0) \ket{0}&=& \int_0^\infty d\kappa^2 
                                   w_{(1)}(\kappa^2)\Delta^{(+)}(x,\kappa^2)\NN
&=& M\Delta^{(+)}(x,M^2)\NN
&=& \frac{M^2}{4\pi^2}\frac{K_1(M\sqrt{-x^2})}{\sqrt{-x^2}} \label{ETresult}
\EQA
for $x^2<0$.
Here we have used the expression for the free fermion spectrum 
 $w_{(1)}(\kappa^2)=M\delta(\kappa^2-M^2)$, and $\Delta^{(+)}(x,\kappa^2)$
 is one of the invariant delta functions.
Restriction of Eq.~(\ref{ETresult}) to the case of equal LF time, i.e.
  $x^+\to 0$ yields ($x^2=-x_\perp^2<0$)
\BQ
\bra{0} \bar\Psi (x) \Psi(0) \ket{0}|_{x^+=0}=
\frac{M^2}{4\pi^2}\frac{K_1(M\sqrt{x_\perp^2})}{\sqrt{x_\perp^2}}.
\label{correct_result}
\EQ
The chiral condensate is obtained as the limit $x_\perp\to 0$.
It is evident that the result depends on the mass $M$ nontrivially.

However, the LF canonical procedure gives only a trivial dependence 
  of $M$, as is seen in Eq.~(\ref{mean_field}).  
The origin of this discrepancy can be traced back to the integral 
  representation of $\Delta^{(+)}(x,M^2)$.
With the LC coordinates, it is  given by 
\BQ
\Delta^{(+)}(x,M^2)=\frac{1}{(2\pi)^3}\int_0^\infty
 \frac{dp^+}{p^+}\int_{-\infty}^\infty d^2p_\perp 
 \ {\rm e}^{-i\big(\frac{p_\perp^2+M^2}{2p^+}x^+
               + p^+x^-- p_\perp^i x_\perp^i\big)} .
\label{integral_rep}
\EQ
Clearly, $p^+=0$ is a singular point, and taking $x^+$ to be nonzero 
  safely regulates this singularity.
Therefore, a nontrivial $M$ dependence remains after the $p^+$ integration,
  as is seen in Eq.~(\ref{correct_result}). 
On the other hand, if we first set $x^+=0$ in the integral, the result 
  becomes (erroneously) independent of $M$.
This corresponds to the result of the naive LF quantization, which is 
  formulated on the surface $x^+=0$. 
We thus see that the $p^+$ integration does not commute with the LC 
  restriction $x^+\to 0$. 
We must be careful when we encounter singularities of 
  the type $1/p^+$, and have to resort to some prescription for treating them 
  in the canonical LF quantization scheme.
This problem was first recognized by Nakanishi and Yamawaki\cite{Naka-Yama}\  
  long ago in the context of the scalar theories 
  [because $\bra{0}\phi(x)\phi(0)\ket{0}=\Delta^{(+)}(x,M^2)$] and recently 
  by ourselves\cite{Ita-Mae,Thesis,Itakura,I,II} in fermionic theories 
  (see also Ref.~\citen{Lenz}). 

To remedy this problem within the framework of the canonical LF quantization, 
   it is quite natural to supply the mass dependence 
   as the regularization of the divergent integral $\int dp^+/p^+$:
\BQA
\bra{0} \bar\Psi (0) \Psi(0) \ket{0}
&=&-\frac{M}{(2\pi)^3}\int_0^\infty \frac{dp^+}{p^+}\int_{-\infty}^{\infty}
d^2p_\perp\NN
&\to & -\frac{M}{(2\pi)^3}\int_0^\infty \frac{dp^+}{p^+}\int_{-\infty}^{\infty}
d^2p_\perp f_\Lambda(p,M),
\EQA
where $f_\Lambda(p,M)$ is a cutoff function for the IR singularity.
Since the prefactor of the integral tells us the mass of the system, 
  we use $M$ in the cutoff function.
 From our point of view, we do not know how to determine the appropriate 
  form of the cutoff function $f_\Lambda(p,M)$.
In fact, there are many possibilities that give the same result 
  if renormalization is done properly.
For example, Eq.~(\ref{integral_rep}) suggests the heat kernel 
  regularization\cite{Ita-Mae}\ 
  (as seen by replacing $x^+$ by the imaginary time $-i\tau$).
Other schemes are the three- or four-momentum cutoff\cite{Heinzl} 
  $|\p|< \Lambda_3,\ 
  p^2<(\Lambda_4)^2$, and the parity invariant 
  cutoff\cite{Ita-Mae,Thesis,Itakura,I,II} $p^\pm <\Lambda_{\rm PI}$, 
  which we used in Sect.~III.
When we rewrite these conditions in terms of the LC variables  
  $p^+$ and $p_\perp$, we necessarily use the dispersion relation.
Thus the mass dependence enters into the cutoff functions $f_\Lambda(p,M)$,
  as we saw in Sect.~IIIB in relation to Eq.~(\ref{caic}).
The parity invariant cutoff is an economical and tractable scheme.
All of these cutoff schemes are related to the symmetries 
  which the system should have.
Cutoff schemes without symmetry considerations could give erroneous results
  even if they contain the mass dependence.\cite{Heinzl_cutoff}

One thing to be noted is the peculiarity of the mode expansion 
  (\ref{mode_exp}).
It is evident that the mode expansion contains no mass dependence.
Essentially, it is this (mass-independent) mode expansion that 
  caused the above encountered difficulties in the chiral condensate.
On the other hand, such independence of the mass, in turn, implies that 
  our mode expansion allows fermions with {\it any value of the mass}.
In other words, the LF vacuum does not distinguish the value of 
  the fermion mass.
Therefore we can regard the vacuum of a massless fermion as that 
  of a massive one.
The mass of the fields is determined by the Hamiltonian.
This is the reason that  we can have the trivial vacuum while having 
  a nonzero fermion condensate.
This phenomenon is not limited to our specific mode expansion
   but is common to all LF field theories.
Indeed, even if we expand a fermion field with free spinor wave functions
  $u(p)$ and $v(p)$, we have no mass dependence.\cite{Heinzl_review} \\

\subsection{Multiple Hamiltonians}

In the previous section, we saw that the gap equations are obtained from 
  the constraint equations characteristic of the LF formalism.
In general, a gap equation for an order parameter allows several independent 
  solutions corresponding to different phases.
This also implies that there are several solutions to one 
  constraint equation.
Selecting a nontrivial value of the order parameter 
  corresponds to selecting a specific solution of the constraint.
If we substitute the solution into the Hamiltonian, 
  the Hamiltonian depends on the solutions we choose.
In other words, we have multiple Hamiltonians corresponding to 
   possible solutions of the constraint.
This was shown explicitly both in scalar models\cite{multiple}\ 
  and in a fermionic model.\cite{II}\ 

Let us explain this point more concretely.
Suppose we have constraints characteristic of the LF formalism
$$
\Omega_i^{\rm LF}[\xi_i, \eta_i]=0,
$$
where the $\xi_i (\eta_i)$ represent independent (dependent) variables.
In the previous examples, the constrained variables $\eta_i$ were
   the bad component of a fermion and the zero modes of the scalar fields.
By solving these equations, we obtain two kinds of solutions, 
  symmetric ones $\eta_i=\eta_i^{\rm sym}[\xi_i]$ and broken ones 
  $\eta_i=\eta_i^{\rm br}[\xi_i]$.
Since the original canonical Hamiltonian is a function of both $\xi_i$ and 
  $\eta_i$, we have different Hamiltonians corresponding to the different 
  solutions: 
\BQA
H_{\rm sym}[\xi_i]&=&H[\xi_i, \eta_i=\eta_i^{\rm sym}],\NN
H_{\rm br}[\xi_i]&=&H[\xi_i, \eta_i=\eta_i^{\rm br}].\nonumber
\EQA
This contrasts sharply with the usual description of symmetry breaking.
In the standard description,  we have multiple vacua but only one Hamiltonian.
However, in our case, the vacuum remains the trivial Fock vacuum,
  but we have multiple Hamiltonians.
Therefore {\it the information concerning the nontrivial vacuum structure 
  in the usual scheme is moved into the Hamiltonian 
  in the LF formalism}.
This is a very unique way of realizing chiral symmetry breaking, 
  and it inevitably causes various unusual situations, as we discuss below.

Once we have the Hamiltonian, we can compute the wavefunctions and 
  mass spectra 
  of excited states by solving the  LF bound state equation.
This is one of the important aspects of the LF Hamiltonian formalism.
For example, mesonic states in the broken phase can be obtained by 
  using a Hamiltonian in the broken phase $H_{\rm br}$ as 
\BQ
H_{\rm br}\ketrm{meson}=\frac{P_\perp^2+M_{\rm meson}^2}{2P^+}\ketrm{meson},
\label{LFBSE}
\EQ
where $\ketrm{meson}$ can be expanded in terms of various Fock states:
$$
\ketrm{meson}=\ket{q\bar q}+\ket{q\bar q g}+\ket{q\bar q q\bar q}+\cdots.
$$
As we discussed in the Introduction, several kinds of nonperturbative 
  techniques are useful for solving the above equation. 
In particular, due to the simplicity of the vacuum, it is natural to 
  expect that lower exited states can be economically described by 
  a few number of the Fock states with only a few constituents.
For example, the Tamm-Dancoff approximation, which truncates the
  Fock state expansion to a few terms, is effective in the LF quantization.

One important problem is how to determine the true Hamiltonian from several 
  possible Hamiltonians.
Since we do not compute the effective potential, it seems that we do not 
  have any criterion to select the correct Hamiltonian.
However, if we use the wrong Hamiltonian in the LF bound state equation, 
  its solutions will include  a tachyonic state with negative mass square.
This can be a rule for eliminating wrong Hamiltonians.
We comment on this in the next subsection in the context of 
  a concrete example.
\\

\subsection{The NG bosons}

Now let us turn other important physical phenomena in the broken phase, 
  those of the NG bosons.
The NG theorem tells us that if a continuous symmetry breaks down 
  spontaneously, there always appears a massless state associated with
  the broken symmetry.
The massless nature of the NG boson is ensured as long as the symmetry 
  is exact at the Lagrangian level. 
However, this is a bad news for the LF formalism.
It is known that it fails to describe a part of the dynamics of massless 
  particles because they can propagate parallel to the $x^+=$ const 
  surface.  
To avoid this, in the previous three examples of Sect.~III, we always worked 
  with explicit symmetry breaking terms: $c\neq 0$ in the sigma model 
  and a nonzero bare mass term in the others.
A small explicit symmetry breaking term gives a small nonzero mass 
  to the NG boson
  and thus cures the situation.
In the following, we show how to obtain the (pseudo) NG boson (with 
  nonzero mass) in the NJL model.
The problems arising from the massless nature of the NG boson are
  commented on below.

As discussed in Sect.~IVB, a mesonic state can be approximated well 
  by a few constituents. 
This is true of a pseudoscalar pionic state in the NJL model,\footnote{
  This is also justified by the $1/N$ expansion.} which should be the 
  NG boson associated with the breaking of chiral symmetry.
Let us consider a pion as the lowest Fock component $\ket{q\bar q}$ 
  with one fermion ($b^\dag$) and one antifermion ($d^\dag$) created on  
  the Fock vacuum
\BQ
\ket{\pi ; \itP}=\frac{1}{\sqrt{N}}\int_0^{P^+}dk^+
  \int_{-\infty}^{\infty}d^2k_\perp \ 
 \Phi^{\alpha\beta}(\k) 
 b_\alpha^{a\dag}(\k)d_\beta^{a\dag}(\itP-\k)\ket{0},
\EQ
where we have used the mode expansion (\ref{mode_exp}), and 
$\Phi^{\alpha\beta}(\k)$ is the LC wavefunction  normalized as 
\BQ
\int_0^1 dx\int\frac{d^2k_\perp}{16\pi^3}\sum_{\alpha\beta}
\vert\Phi^{\alpha\beta}(\k)\vert^2=1.
  \label{norm}
\EQ
We can determine $\Phi^{\alpha\beta}(\k)$ and the pion mass $m_\pi$ 
  by solving the LF bound state equation (\ref{LFBSE}).
Since we are interested in the broken phase, we assume that the coupling is 
  larger than the critical coupling $\lambda>\lambda_{\rm cr}$, and we select 
  the broken phase Hamiltonian $H_{\rm br}$. 
Using the explicit form of the Hamiltonian\cite{II}\ (the leading nontrivial 
  Hamiltonian of ${\cal O}(N^0)$),  we can easily 
  find the LC wavefunction (see Appendix B and Ref.~\citen{II}),
\BQ
\Phi_{\alpha\beta}(\k)= 
  \frac{C_\pi\lambda}{(2\pi)^3}\frac{M}{m}
\frac{\left\{ \left(i k^i_\perp \sigma^i +M\right)\sigma^3 \right\}_{\alpha \beta}}{(k_\perp^2 + M^2) /m_\pi^2 - x(1-x)}
    ~,\label{solpb}
\EQ
where $\k=(x P^+ , k^i_\perp )$ and $C_\pi$ is just a normalization factor.
The Pauli matrix $\sigma^3$ comes from $\gamma_5$ (see our definition 
  (\ref{gammafive})).
The pion mass $m_\pi$ is obtained from a normalization condition. 
If we use the ``extended parity-invariant cutoff''\cite{Bentz,I}
\BQ
\frac{k_\perp^2+M^2}{x}+\frac{k_\perp^2+M^2}{1-x} < 2\Lambda^2, \label{EPI}
\EQ
the pion mass for a small bare fermion mass $m$ is estimated as 
\BQ
 m_\pi^2 = \frac{NZ_\pi}{\lambda M}m + {\cal O}(m^2)~, \label{pion_mass}
\EQ
where $Z_\pi$ is a cutoff dependent factor:
$$
Z_\pi^{-1}= \frac{N}{8\pi^2} 
 \left[ \ln \left(\frac{1+\beta}{1-\beta} \right)
    - 2\beta ~ \right], \quad \beta=\sqrt{1-2M^2/\Lambda^2}. \label{zp}
$$
Obviously $m_\pi$ vanishes in the chiral limit $m\rightarrow 0$, 
 and this state is indeed identified with the NG boson.
It should be noted that $m_\pi=0$ is realized
  even though the pion is described by the lowest Fock component only.
This is due to the exact cancellation of the kinetic energy 
  and the potential energy (see Appendix B).
Also, in the chiral limit, the LC wavefunction becomes independent of $x$,
  which implies that the pionic state looks like a point particle in the 
  longitudinal direction.
This agrees with the results in two-dimensional QCD.

Similarly, we solved the LF bound state equation (\ref{LFBSE}) for 
  the scalar meson state $\sigma$.\cite{II}\ 
In the chiral limit, the mass of $\sigma$ is given by twice  the 
  dynamical fermion mass $m_\sigma=2M$, and the LC wavefunction turns 
  out to have a very narrow peak at $x=1/2$.  
Thus, the scalar meson can be understood as the usual constituent state.

Now that we have the pion LC wavefunction, we can explicitly compute 
  various physical quantities and relations.
In addition to the form factor\cite{Heinzl_review} 
  and the distribution amplitude,\cite{Bentz} which are intimately related to
  the LC wavefunction, we can also calculate the pion decay constant 
  $f_\pi$ and can check the PCAC relation.
The decay constant $f_\pi$ is defined by 
\begin{equation}
 i P^\mu f_\pi = \bra{0} J^\mu _5 (0) \ket{ \pi;  \itP},
\end{equation}
where we use the current (\ref{Chiral_current}) as mentioned above. 
The result is $f_\pi = 2M Z_\pi^{-1/2} + {\cal O}(N^0)$.
Together with the pion mass (\ref{pion_mass}), 
  we verify the Gell-Mann, Oakes and Renner (GOR) relation,
\BQ
 m_\pi^2 f_\pi^2 = -4m \bra{0} \bar \Psi \Psi \ket{0}.\label{GOR}
\EQ

The PCAC relation can also be checked by using the state $\ket{\pi ; \itP}$.
Introducing the pseudoscalar field $ \pi(x) = Z_\pi^{-1/2}
   (\lambda/N)\bar \Psi i \gamma_5 \Psi(x)$ normalized 
  as $ \bra{0} \pi (0) \ket{\pi ; \itP} = 1$, we arrive at the 
  PCAC relation
\BQA
\partial_\mu J^\mu_5 &=& 2m  \bar \Psi i \gamma_5 \Psi  \nonumber \\
 &=& m_\pi^2  f_\pi \pi(x),\label{PCAC}
\EQA
where we have used the usual current (\ref{Chiral_current}) again.

To this point, we have treated the broken phase Hamiltonian 
  $H_{\rm br}$ to solve the 
  LF bound state equation (\ref{LFBSE}).
What happens if we chose the symmetric phase Hamiltonian but with large 
  coupling, $\lambda>\lambda_{\rm cr}$?
Recall that the gap equation has a symmetric solution $M\to 0$ ($m\to 0$)
  even for $\lambda>\lambda_{\rm cr}$.
In this case, if we solve the bound state equation, we have a
  {\it negative mass square}, which is not physically acceptable.\cite{II}\  
We can verify that the mass square is positive only for 
  $\lambda<\lambda_{\rm cr}$, where we have  only one symmetric solution.
Hence, the emergence of such tachyonic states should be a criterion for
  eliminating  unphysical Hamiltonians.   
\\

\subsection{Modified chiral transformation}

Recall the discussion about the order parameters in Sect.~IIB.
There, we argued that the vacuum expectation value of the 
  fermion bilinear operator $\langle \bar\Psi\Psi\rangle$
  is not exactly an order parameter in the usual sense.
This is because the usual treatments involving order parameters are based on 
   the chiral transformation law (\ref{ChiralET}), but if this law  
   held in the LF formalism, it would always give 
   $\langle \bar\Psi\Psi\rangle =0$, due to $Q_5^{\rm LF}\ket{0}=0$.
Since the invariance of the vacuum under the LF chiral transformation
   is a kinematical fact, and thus independent  of the symmetry,  
   the above observation is serious for the description of the broken phase. 
However, this can be resolved again by the broken phase 
  solution of the characteristic constraints.
Let us go back to the discussion in Sect.~IVA.
There, we argued that if we select the nonzero chiral condensate of the 
  gap equation (\ref{lowest_const_NJL}) in the NJL model, the fermionic 
  field becomes a massive free fermion.
This implies that {\it the operator structure of the dependent variables
   changes according to the phases}.
In the chiral Yukawa model, the operator structure of the longitudinal 
  zero modes, and subsequently of the bad spinor component in the broken phase,
  should be different from those of the symmetric phase.
Therefore, the chiral transformation of the dependent variables also becomes
  modified. 
This can be easily understood from the fact that even the free fermion field 
  transforms unusually if it has a nonzero mass: 
\BQ
\left[Q_5^{\rm LF}, \Psi_M\right]=\gamma_5 \Psi_M - \sqrt2 M \gamma_5 \frac{1}{i\del_-}\psi ,
\EQ
where $\Psi_M$ is the massive free fermion operator given 
  by Eq.~(\ref{massive_fermion}).
Similarly, in the NJL model, the fermion field acquires a dynamical mass 
  in the broken phase, and its transformation is necessarily changed.
This ``modified chiral transformation'' in the broken phase propagates
  into the relation (\ref{ChiralET}).
Indeed,  explicit calculation in the chiral limit of the NJL model\cite{II} 
  (and also of the chiral Yukawa model\cite{I}) yields
  the modified transformation law
\BQ
\left[ Q_5^{\rm LF}, \bar \Psi i \gamma_5 \Psi (x) \right] 
   = - 2i \bar \Psi\Psi(x) + M \Delta(x), \label{mod_chiral}
\EQ
where the extra term $M \Delta(x)$ is a function of the fermion operator.
This unusual chiral transformation, however, is consistent with 
   $Q_5^{\rm LF}\ket{0}=0$, because 
   $- 2i \bra{0}\bar \Psi\Psi(x)\ket{0}
 + M \bra{0}\Delta(x)\ket{0}=0$. 
Conversely, as long as the vacuum is invariant under the LF chiral 
  transformation, the ordinary transformation law (\ref{ChiralET}) must be 
  modified as in Eq.~(\ref{mod_chiral}).
Note also that Eq.~(\ref{mod_chiral}) implies that if we set $M=0$ 
  (the symmetric phase), the usual transformation law is recovered.

The same arguments hold for the sigma model\cite{Tsujimaru-Yamawaki} 
  (and also for the chiral Yukawa model\cite{I}).
The transformation law of the scalar zero mode
   $\pi_0(x_\perp)$ is given by  
\BQ
\left[ Q_5^{\rm LF}, \pi_0(x_\perp) \right]=- 2i \sigma_0(x_\perp) + 
  \Delta \pi_0(x_\perp),
\EQ     
where $\Delta \pi_0$ is an extra term that does not appear in the usual 
  transformation law.

All the operators including dependent variables should be transformed 
  unusually in the broken phase under the LF chiral transformation.
This is one of the inevitable consequences of the coexistence of 
  the symmetry breaking and the trivial vacuum.
Since we cannot compute the transformation of the dependent variables until 
  solving the constraint equations, the extra terms  inevitably become 
  model dependent.  
This prevents us from giving generic discussion about the broken phase 
  in the LF formalism.
For example, it is very difficult to prove the NG theorem in general.
However, we can say that  the vacuum expectation value of the model-dependent 
 extra term should always 
  coincide with that of the order parameters 
  to ensure trivial relations such as 
  $\bra{0}[ Q_5^{\rm LF}, \bar \Psi i \gamma_5 \Psi (x) ] \ket{0}=0$
  and $\bra{0}[ Q_5^{\rm LF}, \pi_0(x_\perp) ]\ket{0}=0$.  
\\

\subsection{Nonconservation of the LF chiral charge}

The same arguments as those given in Sect.~IVD hold for the Hamiltonian.
This was already seen partly in Sect.~IVB.
The new concept of multiple Hamiltonians  is a direct consequence of the
  fact that the Hamiltonian contains dependent variables. 
It is natural to expect that the chiral transformation of the Hamiltonian 
   is also modified in the broken phase.
The Hamiltonian is not invariant under the LF 
  chiral transformation in the broken phase.
In fact, in the NJL model,\cite{II}\  
  we find the commutator $[Q_5^{\rm LF}, H ]$ is really nonzero in the
  broken phase, while zero in the symmetric phase: 
\BQA
&&\left[Q_5^{\rm LF} , H_{\rm br} \right]  = M \Delta H_{\rm br}\neq 0,
     \label{nonconserv_H}\\
&&\left[Q_5^{\rm LF} , H_{\rm sym} \right]=0.
\EQA
In other words,  the null-plane chiral charge 
  $Q_5^{\rm LF}$ is not conserved in the broken phase.
This has been pointed out by several people as a characteristic 
  feature of the chiral symmetry breaking on the 
  LF.~\cite{nonconserv,Tsujimaru-Yamawaki}

The result (\ref{nonconserv_H}) should be consistent with the current 
  divergence relation Eq.~(\ref{current_div}).
Integrating it over space, we have
\BQ
  \partial_+ Q_5^{\rm LF} = \frac{1}{i} \left[ Q_5^{\rm LF} , H_{\rm br} 
  \right] 
                          = 2m \int dx^- d^2 x_\perp 
                                 \bar\Psi i \gamma_5 \Psi ~.
  \label{nonconserv}
\EQ
Therefore if the LF chiral charge is not conserved in the chiral limit, 
  the integral on the right-hand side  must exhibit the singular behavior 
\BQ
\int dx^- d^2 x_\perp \bar \Psi i \gamma_5 \Psi \propto \frac{1}{m}~.
\label{singular}
\EQ
By using the solution of the fermionic constraint, we can directly 
  verify this singular behavior and the extra term in Eq.~(\ref{nonconserv_H}).

The importance of such singular behavior for making the NG boson 
  meaningful has been stressed by Tsujimaru et al. in the context of 
  scalar theories.~\cite{Tsujimaru-Yamawaki}\ 
Assuming the PCAC relation, they showed that the zero mode of the 
  NG boson has a singularity  $\sim m_{\rm NG}^{-2}$, where 
  $m_{\rm NG}$ is an explicit symmetry-breaking mass.
If we set $m=0\ (m_{\rm NG}=0)$ from the beginning, the NG boson decouples 
  from the system, and the NG phase is not realized.
In our case, due to the GOR relation (\ref{GOR}) and the PCAC relation 
 (\ref{PCAC}),  we can rewrite Eq.~(\ref{singular}) as
\BQ
\int dx^- d^2 x_\perp \pi_0 (x) \propto \frac{1}{m_\pi^2},
\EQ
which is the same form as that discussed in Ref.~\citen{Tsujimaru-Yamawaki}.
\\

\section{Coupling to gauge fields}

To this point, we have discussed scalar and fermionic fields.
What occurs if we couple gauge fields to them?
As we  see below, this addition has a drastic effect, and the problem
  becomes greatly complicated.
We discuss here the simplest example of the Abelian fields 
  and give a brief comment on the non-Abelian case.

Using the complex scalar field $\phi = ( \sigma + i \pi )/\sqrt2, $
the $U(1)$ gauged sigma model can be equivalently rewritten as
  the Abelian Higgs model, 
\BQ
  {\cal L}= -\frac{1}{4}F_{\mu \nu}F^{\mu\nu} +
            |D_{\mu}\phi|^2 -V(\phi), \quad 
  V(\phi) = \frac{\lambda}{4}({\phi}^*\phi)^2 - \frac{\mu^2}{2}{\phi}^*\phi,
\label{AH}
\EQ
where $D_{\mu} =\partial_\mu -ieA_\mu$ and 
  $F_{\mu\nu}=\del_\mu A_\nu -\del_\nu A_\mu$.
This model exhibits the Higgs mechanism, where the global $U(1)$ 
  symmetry (which corresponds to the chiral symmetry in the sigma model) 
  breaks down spontaneously and the photon acquires a nonzero mass.
No NG boson appears because it is absorbed into the photon.

In the following, we consider 1+1 dimensions for simplicity, 
  but it is straightforward to generalize our discussion to 
  3+1 dimensions.\cite{AH}\  
As we have scalar fields, we use the DLCQ method.
Periodic boundary conditions are imposed on the gauge fields 
  as well as the scalar field.
Thus, the zero modes we treat are a scalar zero mode $\phi_0$ and 
  a gauge field zero mode $\zm{A}_-=\int_{-L}^L dx^- A_-(x)/2L$. 
In the Abelian Higgs model, the canonical structure of the zero mode 
  is different from that of the sigma model.\cite{AH,Topology}\  
To see this, let us consider the Euler-Lagrange  equation of the Higgs field,
\BQ
2D_-D_+\phi -ie\phi\ \Pi^-+ \frac{\del V}{\del \phi^*}=0,\label{EL}
\EQ
where $\Pi^-=F_{+-}=\del_+A_--\del_-A_+$ is the momentum conjugate to $A_-$.
The covariant derivative $D_-$  can be replaced by a normal derivative 
$\del_-$ if we introduce the (LF) spatial Wilson line
\BQ
W(x^+,x^-) 
\equiv \exp \left\{ie\int_{-L}^{x^-}dy^-A_-(x^+,y^-)\right\}.
\EQ
Equation (\ref{EL}) can  then equivalently be rewritten as 
\BQ
\del_-[2W^{-1}D_+\phi]
=W^{-1}\left[
ie\phi\ \Pi^--\frac{\del V}{\del\phi^*}\right]\equiv \omega .\label{EL2}
\EQ
Unlike the case for the previous scalar model (see Sect.~IIIA.), the 
   integral of the 
  left-hand side of this equation does not necessarily 
  vanish, because the Wilson line is not periodic in general.
This means that {\it only if the Wilson line is periodic}, i.e. $W(-L)=W(L)$, 
  does the $x^-$  integration of Eq.~(\ref{EL2}) generate a constraint 
  analogous to the usual zero-mode constraint. 
If we employ the LC axial gauge, the unfixed gauge degree of freedom 
  turns out to be $\zm{A}_-$ only. 
The Wilson line is periodic when  $\zm{A}_-$ takes the discrete values 
  $\zm{A}_-=\pi N/eL,\ N\in {\bf Z}$. 
These discrete points correspond to the zero modes of the covariant 
  derivative $D_-=\del_--ie\zm{A}_-$.
(Recall that the usual zero-mode constraint is related to the zero mode 
  of the derivative operator $\del_-$.) 
Gauge fields with different $N$ are equivalent, because they are related 
  by the large gauge transformation, which is associated with the nontrivial 
  homotopy group $\pi_1(S^1)=\Z$.
Therefore the canonical structure of the model is summarized as follows:\\

\begin{description}
\item{(A) Periodic Wilson line}: \ 
$W(-L)=W(L)\ \Longleftrightarrow\ \zm{A}_-=\pi N/eL$.\\
  Here there is the extra constraint $\int_{-L}^L dx^- \omega =0$.
  One mode in the scalar field is a constrained variable.
  
\item{(B) Non-periodic Wilson line}:\  
$W(-L)\neq W(L)\ \Longleftrightarrow\ \zm{A}_- \neq \pi N/eL$.\\
Here there is no extra constraint, and all the modes of the scalar field are 
  dynamical. 
The gauge field zero mode $\zm{A}_-$ is also dynamical.\\
\end{description}

\noindent 
For case (A), the same arguments as those given in the case of the 
  sigma model are applicable. 
We can solve the constraint, and the nonperturbative (tree level) solution  
  gives a nonzero vacuum expectation value of $\phi$.
However, this is an exceptional point in the phase space,  
  and we have to treat case (B) in general.
In this case, we do not have the zero-mode constraint, but there is no problem.
Our situation is rather similar to that for the usual ET calculation. 
We only need to evaluate the vacuum energy and find the true vacuum, as usual.
It is easy to find classical configurations that minimize the LF
  energy $P^-$.  
The LF energy becomes zero if and only if the field configuration is given by 
\BQ
\phi = \frac{\mu}{\sqrt{\lambda}}\ {\rm e}^{i\pi N\frac{x^-}{L}},\quad 
\zm{A}_-= \pi N/eL.\quad (N \in {\bf Z})\label{config}
\EQ
This configuration gives a nonzero vacuum expectation value to the scalar 
  field.
Note that Eq.~(\ref{config}) should be classified as case (A). 
Indeed, the scalar field $\phi$ above is related to the zero-mode solution 
 $\phi_0=\mu/\sqrt{\lambda}$ via the large gauge transformation.
We can formulate the quantum problem in case (B) by expanding the 
  fields around this configuration.\cite{Topology}\ 
Since the zero modes are all dynamical, 
  the true vacuum is not the trivial Fock vacuum, but should be determined 
  as the lowest energy eigenstate of the Schr\"odinger equation within 
  the zero-mode sector.\footnote{Furthermore, we can explicitly construct 
  the $\theta$ vacua by imposing on the vacuum invariance under the 
  large gauge transformation.\cite{Topology} }

Hence, the canonical structure in the zero-mode sector is not so simple,  
  due to the presence of the gauge field.
Depending on the periodicity of the Wilson line, we have to consider two 
  cases separately, but it appears that these two cases are connected,  
  and we can describe the symmetry breaking (the Higgs mechanism)
  in both cases.

The same structure is observed in non-Abelian gauge theories.
For example, in the 3+1 dimensional $SU(2)$ Yang-Mills theory,\cite{Franke}\  
 at the points where the gauge field zero mode takes the values 
  $\zm{A}_-\!\!\!{}^a=\delta^{a3} n\pi/gL$ (corresponding to 
  case (A) above), there exist two constraint relations,
\BQ
\int_{-L}^{L}dx^-e^{\frac{i\pi n}{L}x^-}(G^1+iG^2)=0,\label{YM}
\EQ
where $G^a = (D_iF_{-i})^a$ or $G^a=(D_iF_{+-}+D_jF_{ji})^a$ for $a=1,2$
  and $F_{\mu\nu}=\del_\mu A_\nu -\del_\nu A_\mu - ig [A_\mu, A_\nu]$.
Noting that these originate from the zero eigenvalue problem 
  of the covariant derivative $D_-$, we can easily imagine that the same 
  kind of constraint relations will be present in any gauge theory,
  as well as in QCD.
In Ref.~\citen{Franke}, quantum theory in the case with the 
  constraints (\ref{YM}) was not developed due to its complexity.
The physical consequences of these constraints are still not known.
As we saw in the Abelian Higgs model, however, the zero-mode 
  constraint in case (A) has significant meaning, though 
  it appears  as a special case.
Therefore it is interesting to investigate the consequence of 
  the constraints (\ref{YM}) and to understand the zero-mode structure of 
  the gauge theories.
\\

\section{Conclusions}

We have studied how to describe chiral symmetry breaking on the LF. 
Throughout the analyses, the most important point was to clarify the role 
  of the characteristic constraint equations which are present only 
  in the LF formalism. 
Many problems are resolved by carefully treating the constraint equations
  and their solutions. 
We have given clear answers to the three questions posed in the Introduction.
Let us recapitulate them below as our conclusions. \\

\begin{description}
\item {(i)} 
The chiral transformation defined on the LF is different in general from  
  the ordinary chiral transformation.
This is because  half of the fermion degrees of freedom (the bad component) 
  are not independent, and we define the chiral transformation only on the 
  independent degrees of freedom (the good component). 
The difference is evident in the free fermion theory, where the LF chiral 
  symmetry is exact, even with the mass term.
Nevertheless, as far as the models we studied in the present paper 
  are concerned, such extra symmetry does not exist.
This was explicitly checked by solving the constraint equations classically.
If and only if we do not have a bare mass term, the LF chiral 
  transformation is the symmetry of the system, and it is equivalent to 
  the ordinary chiral symmetry. 
(Sects.~IIB and III.)
 
\item {(ii)}
We have seen in several models that the zero-mode constraint and 
  the fermionic constraint play the same role.
The gap equations result from these constraint equations. 
This is natural, because the constraints were originally a part of the 
  Euler-Lagrange equations and should carry the information concerning
  the dynamics.
However, technically this is not evident, and we have to be careful about 
  the infrared divergence of the longitudinal momentum integration.
This is related to the fundamental problem in the LF formalism, 
  the loss of the correct mass dependence from two-point functions.
Without the proper treatment of the IR divergence, we cannot obtain  meaningful
  gap equations.
(Sects.~III and IVA.)

\item {(iii)}
As a result of the coexistence of a nonzero chiral condensate and the trivial 
  Fock vacuum, many unusual consequences follows.
Most of them can be properly understood if we carefully treat the 
  characteristic constraint equations.

\begin{description}
\item {(iii-a)}  
Even if we select a nontrivial solution to the gap equation, the vacuum 
  still remains the Fock vacuum.
In general, there are several independent solutions to the gap equation, 
  which implies several solutions to the constraint equations. 
Consequently, we obtain multiple Hamiltonians, but with one trivial Fock 
  vacuum.
Multiple vacua in the ET formalism corresponds to multiple Hamiltonians in the
  LF formalism. 
Effects of the usual nontrivial vacuum physics are converted into the 
  Hamiltonian in the LF formalism.
(Sect.~IVB.)

\item {(iii-b)}
Since we have a trivial vacuum, even the NG boson can be described by a few 
  constituents. 
In the NJL model, we explicitly constructed  the pionic state and obtained the 
  LC wavefunction and the mass.  
Using the LC wavefunction, we can directly calculate various physical 
  quantities (e.g., the pion decay constant) and relations 
  (e.g., the GOR and PCAC relations).
(Sect.~IVC.)

\item {(iii-c)}
Since the null-plane charge always annihilates the vacuum, it appears that 
  a nonzero chiral condensate is inconsistent with the chiral transformation 
  law of $\bar\Psi i\gamma_5\Psi$ (see Eq.~(\ref{ChiralET})). 
However, in the broken phase, the chiral transformation of the dependent 
  variables is necessarily changed, and, subsequently, the 
  transformation law for $\bar\Psi i\gamma_5\Psi$ is also modified, 
  so that there is no inconsistency in the equation (Eq.~(\ref{mod_chiral})).
(Sect.~IVD.)
 
\item {(iii-d)} 
The Hamiltonian in the broken phase does not commute with the LF chiral charge.
In other words, the LF chiral charge is not conserved in the broken phase.
This is essentially because the Hamiltonian contains dependent variables and
  they are transformed unusually in the broken phase under the LF chiral 
  transformation.
This is also related to the singular behavior of the spatial integration 
  of the NG field, which appears in the limit of small explicit symmetry
  breaking mass.
(Sect.~IVE.)

\end{description}
\end{description}

\noindent Therefore, as far as we are concerned with scalar and fermionic
   systems, the dynamical chiral symmetry breaking can be satisfactorily 
   described by the LF formalism. 

The next problem to be addressed, of course, regards the effects of gauge 
  fields. 
In the presence of the gauge fields, the canonical structure of the 
  zero-mode sector becomes greatly complicated. 
In particular, the scalar and gauge-field zero modes remain  dynamical
  in almost all of phase space.
The physics of the zero mode is still important for describing symmetry 
  breaking, and we indeed demonstrated this in the Abelian Higgs model 
 (Sect.~V). 
For chiral symmetry breaking in QCD, we need to further 
  understand the physical implications of the gauge field zero modes.

\section*{Acknowledgements}
We would like to thank 
  W. Bentz, K. Harada, T. Heinzl, M. Tachibana, S. Tsujimaru, 
  K. Yamawaki, and K. Yazaki for various discussions. 
We are also grateful to V.~A. Franke, T. Hatsuda, T. Matsui, H. Toki and 
  D. Vautherin for their encouragements and interest in our work.
\\

\appendix

\setcounter{equation}{0}
\section{Conventions}

We follow  the Kogut-Soper convention.\cite{Review}\ 
First of all, the LF coordinates are defined as
\begin{equation}
x^\pm = \frac{1}{\sqrt2}(x^0\pm x^3), \quad
x^i_\perp = x^i, \quad(i=1,2) 
\end{equation}
where we treat $x^+$ as ``time''. 
The spatial coordinates $x^-$ and $x_\perp$ 
  are called the longitudinal and transverse 
  directions, respectively. 
Therefore, our metric is 
\BQ
g_{\mu\nu}=\left(\matrix{
0&1&0&0\cr
1&0&0&0\cr
0&0&-1&0\cr
0&0&0&-1\cr
}\right)
\EQ
for $\mu,\nu=(+,-,1,2)$.
The inner product between two four vectors is given by
\BQA
p\cdot x&=&p_+x^+ + p_-x^- + p_{\perp i} x_\perp^i\NN
        &=&p^-x^+ + p^+x^- - p_{\perp}^i x_\perp^i.
\EQA
 From this, we can find that $p^-=(p^0-p^3)/\sqrt2$ is the LF energy 
   and $p^+=(p^0+p^3)/\sqrt2$ is the longitudinal momentum.
Derivatives in terms of $x^\pm$ are defined by
$\partial_\pm=\partial/\partial x^\pm$.
The inverse of $i\del_-$ is defined by 
\BQ
\frac{1}{i\del_-}f(x^-)=\int_{-\infty}^\infty dy^- 
\frac{\epsilon(x^--y^-)}{2i} f(y^-),
\EQ
where $\epsilon(x)$ is a sign function. 
The operator $\del^2$ is given by $\del^2=2\del_+\del_- - \del_\perp^2$.

It is useful to introduce projection operators $\Lambda_\pm$ 
  defined by
\begin{equation}
\Lambda_{\pm}=\frac12 \gamma^\mp \gamma^\pm
=\frac{1}{\sqrt2}\gamma^0\gamma^\pm.
\end{equation}
Indeed $\Lambda_\pm$ satisfy the projection properties 
$\Lambda_\pm^2=\Lambda_\pm,\ \Lambda_++\Lambda_-=1$, etc.
Splitting the fermion field by use of the projectors as
\begin{equation}
  \Psi = \psi_+ + \psi_- , \quad 
  \psi_{\pm} \equiv \Lambda_{\pm}\Psi ,
\end{equation}
we find that for any fermion on the LF, 
  the $\psi_-$ component is a dependent degree of freedom. 
$\psi_+$ and $\psi_-$ are called the ``good component'' and 
  the ``bad component'', respectively.

As is commented in the text, for practical calculation, we use 
  the two-component representation for the gamma matrices.
The two-component representation is characterized by a specific 
  form of the projectors $(\ref{two-compo})$. 
Then the projected fermions $\psi_\pm$ have only two components.
There are many possibilities that realize Eq.~$(\ref{two-compo})$.
For example, the specific representation 
\BQ
\gamma^0=\pmatrix{
0 & -i\cr
i&0
},\quad
\gamma^3=
\pmatrix{
0& i\cr
i&0
}, \quad
\gamma^i=
\pmatrix{
-i\sigma^i&0\cr
0&i\sigma^i
}
\EQ
is used in Ref.~\citen{Hari-Zhang}.
In this paper, however, we choose the representation given in 
  Eq.~(\ref{two-compo-gamma}). 
This representation allows us to easily extract results of 
  the 1+1 dimensions from those of the 3+1 dimensions.
\\

\section{LC Wavefunction of a Pion in the NJL Model}

Here we discuss further the LC wavefunction of a pion in the LF NJL model. 
First, using the explicit form of the Hamiltonian 
  (Eqs.~(3.29)--(3.32) in Ref.~\citen{II}),  we can 
  find the spinor structure of a pseudoscalar state  should be 
\BQ
\Phi_{\alpha\beta}(\k)= 
 \left\{ \left(i k^i_\perp \sigma^i +M\right)\sigma^3 \right\}_{\alpha \beta}
 \phi_\pi (x, k^i _\perp ),\label{solp1}
\EQ
where $\k=(x P^+ , k^i_\perp )$.
The Pauli matrix $\sigma^3$ which comes from $\gamma_5$ ensures that this 
   is a pseudoscalar state.
The spinor independent part $\phi_\pi (x,k_\perp^i)$ satisfies  
  the integral equation
\BQ
m_\pi^2 \phi_\pi (x, k_\perp^i)
= \frac{k_\perp^2+M^2}{x(1-x)} \phi_\pi (x , k_\perp^i)         
 -  \frac{\lambda \alpha}{(2\pi)^3}\frac{1}{x(1-x)}
    \int_0^1 dy  \int d^2 \ell_\perp 
      \frac{\ell_\perp^2 + M^2}{y(1-y)}
    \phi_\pi (y , \ell_\perp^i) ~, \label{LFBSE_pion}
\EQ
where $\alpha$ is just a numerical factor:
$\alpha^{-1}=m/M + 2 \lambda/(2\pi)^3 \cdot \int d^2 q_\perp \int_0^1 dx/x $.
Its solution is easily found as
\BQ
\phi_\pi (x, k_\perp^i)
   =C_\pi \frac{\lambda}{(2\pi)^3}\frac{M}{m}
    \frac{1}{(k_\perp^2 + M^2) /m_\pi^2 -x(1-x)}~,\label{solp2}
\EQ
where $C_\pi$ is a normalization constant: 
\BQ
C_{\pi}=\int_0^1dx\int d^2k_\perp \phi_{\pi}(x,k_\perp^i). \label{Cpi}
\EQ
Combining Eqs.~(\ref{solp1}) and (\ref{solp2}), we obtain Eq.~(\ref{solpb}).
Equation~(\ref{Cpi}) also leads to an equation for $m_\pi$ (Actually 
$C_\pi$ is determined by the normalization condition Eq.~(\ref{norm}))
\BQ
\frac{1}{\lambda}= \frac{M}{m}\int_{0}^{1} dx \int\frac{d^2 k_\perp}{(2\pi)^3}
     \frac{m_\pi^2}{ k_\perp^2 + M^2 - m_\pi^2 x(1-x) }.
\EQ
This nonlinear equation can be analytically solved for small bare mass case.
As is shown in the text, if we introduces the cutoff (\ref{EPI}), 
we find Eq.~(\ref{pion_mass}).

In Eq.~(\ref{LFBSE_pion}), the first term corresponds to the
   kinetic energy part of the fermion and antifermion, with 
  constituent mass $M$, and the 
  second term corresponds to the potential energy part.
The constituent picture for the massless pion is realized by an exact 
  cancellation of the kinetic and potential terms.
Note that the kinetic term is just a summation of the kinetic energies of 
  the fermion and antifermion  with mass $M$.\\

\end{document}